\documentclass{aa}  

\usepackage{graphicx}
\usepackage{txfonts}
\usepackage{lipsum}
\usepackage{subcaption}         
\usepackage{lscape}             
\usepackage{placeins}           
\usepackage{gensymb}
\usepackage{multirow}
\usepackage{color}
\usepackage{rotating}
\usepackage{makecell}

\newcommand\ions[2]{{#1}\,{\sc #2}} 
               
\begin{document}


   \title{Chemical signatures of planetary systems in their host stars}
     \subtitle{Near-infrared spectroscopy of four planet-hosting wide binaries}

%

   \author{Dongwook Lim\inst{1} 
        \and Sol Yun\inst{2} 
        \and Andreas J. Koch-Hansen\inst{3} 
        \and Sang-Hyun Chun\inst{4} 
        \and Young Sun Lee\inst{5} 
        \and Young-Wook Lee\inst{1} 
        }
        
   \authorrunning{D. Lim et al.}
   \titlerunning{}

   \institute{
   Center for Galaxy Evolution Research \& Department of Astronomy, Yonsei University, 50 Yonsei-ro, Seoul 03722, Republic of Korea,
    \email{dwlim@yonsei.ac.kr}
    \and Department of Earth, Environmental \& Space Sciences, Chungnam National University, 99 Daehak-ro, Daejeon 34134, Republic of Korea
    \and Zentrum f\"ur Astronomie der Universit\"at Heidelberg, Astronomisches Rechen-Institut, M\"onchhofstr. 12-14, 69120 Heidelberg, Germany
    \and Korea Astronomy and Space Science Institute, 776 Daedeokdae-ro, Yuseong-gu, Daejeon 34055, Republic of Korea
    \and Department of Astronomy and Space Science, Chungnam National University, 99 Daehak-ro, Daejeon 34134, Republic of Korea
    }

   \date{Accepted May 7, 2026}

  \abstract
  {An important open question in exoplanet studies is whether planets can leave detectable chemical fingerprints on their host stars.
  While several studies have suggested possible planetary chemical signatures in planet-hosting stars, including the Sun, their true origin remains debated because of stellar birth conditions and evolutionary effects.
  Wide binaries, whose components share a common formation environment and have evolved without significant mutual interactions, provide an ideal testbed for identifying planetary chemical signatures.}
  {Such signatures are often characterized by differential abundance trends with condensation temperature ($T_{\rm cond}$), which traces the partitioning between gaseous and rocky planetary material.
  We aim to investigate whether these trends are associated with the known planetary architectures of planet-hosting wide binaries.}
  {We obtained high-resolution near-infrared spectra of four planet-hosting wide binaries with IGRINS at Gemini-South.
  We measured chemical abundances for both components and analyzed differential abundances in each system.
  We also compiled literature measurements for planet-hosting and non-planet-hosting wide binaries and compared their $T_{\rm cond}$ trends.}
  {Two systems (WASP-160~A/B and WASP-127/TYC~4916-897-1) exhibit statistically significant abundance trends with $T_{\rm cond}$, while HD~20782/HD~20781 shows a weaker correlation and K2-54/K2-54~B is consistent with a flat relation.
  The observed trends display diverse behaviors, including both volatile- and refractory-enhanced patterns in planet-hosting stars.
  Literature comparisons indicate that extreme $T_{\rm cond}$ slopes may occur more frequently among planet-hosting wide binaries, particularly at large binary separations, although the broader statistical picture remains limited by sample size and sample definition.}  
  {Our results indicate that chemical signatures in planet-hosting wide binaries are not universal but vary across systems.
  While planetary architectures may be associated with some of the observed host-star abundance patterns, multiple competing processes are likely to contribute to the observed trends.
  Larger samples of both planet-hosting and non-planet-hosting wide binaries, including those observed with near-infrared spectroscopy, will be essential for more robustly disentangling planetary signatures from stellar and binary effects.}
  

   \maketitle

\nolinenumbers
\section{Introduction} \label{sec:intro}

Our exploration of planetary systems naturally begins with the Sun, the nearest star known to host multiple planets, including the Earth.
Analogous to the intimate physical and chemical link between the Sun and the Earth, understanding the properties of planet-hosting stars is fundamental to exoplanet studies.
As the number of known exoplanets continues to grow, increasing attention has been devoted to characterizing their host stars.
Most previous studies have focused on how stellar properties influence planet formation and evolution \citep{Santos2004, Fischer2005, Adibekyan2012, Buchhave2012, Sousa2021, Yun2024}.
An important but complementary question is whether planets can, in turn, leave detectable chemical signatures in the atmospheres of their host stars.
We note that although direct characterization of exoplanet atmospheres is becoming increasingly feasible with current and upcoming facilities \citep[e.g.,][]{Kaltenegger2017, vanSluijs2025}, stellar spectra remain more readily accessible and provide a powerful indirect probe of planet–star interactions.
This idea was first motivated by studies of the Solar System.
\citet{Melendez2009} reported that the solar chemical abundance pattern differs from those of nearby solar twins, showing a relative depletion of refractory elements \citep[see also][]{Ramirez2009, Bedell2018}.
As refractory elements constitute the primary building blocks of terrestrial planets, \citet{Melendez2009} proposed that this depletion may be linked to the formation of rocky planets in the Solar System.
However, this interpretation remains debated, as alternative explanations unrelated to rocky planet formation have also been proposed \citep[e.g.,][]{GonzalezHernandez2013, Booth2020, Nibauer2021, Carlos2025}.

Current advances in high-precision chemical abundance measurements allow this question to be extended beyond the Solar System to exoplanets and their host stars.
Nevertheless, despite the confirmation of more than 6000 exoplanets to date, identifying chemical fingerprints of planets in host-star atmospheres remains a challenging task.
One major obstacle is Galactic chemical evolution, as stars form in different environments and follow distinct evolutionary pathways, imprinting intrinsic abundance variations that are unrelated to planet formation \citep{Spina2016, Pignatari2023}.
Such effects can obscure or mimic planetary chemical signatures, complicating their interpretation \citep{Adibekyan2014, Nissen2015, Cowley2022}.
In this context, wide binary systems offer a powerful laboratory for isolating potential planetary signatures, as the two stellar components are expected to share a common formation environment and evolutionary history.

Wide binaries are systems with separations ranging from a few astronomical units (au) to several thousand au.
Although various formation scenarios have been proposed \citep[e.g.,][]{Kouwenhoven2010, Reipurth2012, Tokovinin2017, Lee2017}, wide binary components are generally expected to be co-natal and co-eval, and this expectation is supported by their observed chemical homogeneity \citep{Desidera2004, Hawkins2020, Lim2024}.
As a result, wide binaries, which are less affected by strong dynamical interactions than closer binaries and are therefore more likely to host long-term stable planetary systems \citep{Holman1999, Kane2013, Kane2025}, provide a unique and controlled testbed for investigating planetary chemical signatures in host stars, particularly in systems where the two components host different planetary architectures or where only one component is known to host planets.
In such systems, differential chemical signatures can more directly reflect the presence or absence of planets.
Motivated by these advantages, a number of high-resolution spectroscopic studies have targeted planet-hosting wide binary systems \citep[e.g.,][]{Ramirez2015, Saffe2017, Jofre2021, Ryabchikova2022, YanaGalarza2024}.
A key observational approach in these studies is to measure precise differential abundances between the two stellar components, focusing on volatile and refractory elements, which predominantly constitute gaseous and rocky planets, respectively.
In particular, planetary chemical fingerprints are commonly investigated through trends in differential abundances as a function of elemental condensation temperature, following the methodology originally developed for the Sun and solar twins.

Previous studies on planet-hosting wide binaries have suggested that multiple physical processes may imprint planetary signatures on host stellar atmospheres.
One of the earliest proposed scenarios is the depletion of refractory elements in host stars through the sequestration of refractory material into terrestrial planets, as originally suggested for the Sun \citep{Melendez2009, Ramirez2014}.
Subsequently, alternative explanations have been proposed, including the formation of pressure traps associated with giant planet formation, which may inhibit the inward drift and accretion of refractory-rich materials onto the host star \citep{Booth2020, Huhn2023}.
Conversely, an enhancement of refractory elements relative to volatiles may arise from the accretion or ingestion of rocky material triggered by the inward migration of giant planets \citep{Schuler2011, Raymond2011, Mustill2015}.
Planetary engulfment events have also been invoked to explain pronounced refractory enhancements observed in some systems \citep{Oh2018, Nagar2020}.
In contrast, abundance trends with condensation temperature may also originate from processes unrelated to planetary systems, such as differential atomic diffusion or primordial chemical inhomogeneities between stars.
For example, wide binaries formed from larger and less chemically homogeneous gas clouds may show enhanced abundance differences between their components with increasing separation \citep{Ramirez2019}, while differences in atmospheric parameters can lead to differential atomic diffusion that produces abundance differences in stellar atmospheres \citep[e.g.,][]{Liu2021}.
In addition, poorly measured abundances, especially for elements that anchor such trends, can also generate apparently significant patterns even without planet engulfment or other planetary effects \citep[e.g.,][]{Behmard2023}.
Despite detailed modeling of individual systems, disentangling these mechanisms remains challenging because different processes can produce similar abundance patterns, and the inferred chemical signatures are not always consistent with models of known planetary systems \citep[e.g.,][]{Jofre2025}.

In this study, we investigate four wide binary systems in which one or both components host at least one detected planet: HD~20782/HD~20781, WASP-160~A/WASP-160~B, K2-54/Gaia~DR3~2594546351459477632, and WASP-127/TYC~4916-897-1.
Among these, the first two systems have previously been examined using optical high-resolution spectroscopy \citep{Mack2014, Jofre2021}, whereas the latter two have not yet been studied in detail.
In contrast to most previous optical studies of planet-hosting wide binaries, our analysis is based on high-resolution near-infrared (NIR) spectroscopy obtained with IGRINS \citep{Mace2018}.
A key advantage of NIR observations is the improved determination of volatile elements such as C, N, and O from numerous molecular features \citep[see][]{Lim2024, Lim2025}, providing robust low-condensation-temperature anchors that are often difficult to obtain in the optical.
Our first goal is to investigate the element-by-element differential abundance patterns in the four selected systems using high-resolution NIR spectroscopy.
Our second goal is to explore, in a broader statistical sense, whether such chemical patterns show any observational association with currently known planetary architectures in the available wide-binary sample, thereby providing empirical constraints on possible planetary imprints on their host stars.

This paper is organized as follows.
Section~\ref{sec:obs_reduction} describes the observations and data reduction procedures.
Section~\ref{sec:spec} outlines the spectroscopic analysis and chemical abundance measurements.
The resulting chemical properties and their relation to planetary systems are presented in Sections~\ref{sec:result} and \ref{sec:result2}.
In Section~\ref{sec:liter}, we examine planetary effects on host stars using a compiled sample of planet-hosting and non-planet-hosting wide binaries from the literature.
Finally, we summarize our results and discuss the limitations and future prospects of this study in Section~\ref{sec:discussion}.


\section{Observation and data reduction} \label{sec:obs_reduction}

\subsection{Target selection} \label{sec:sub:target}
To investigate the effect of planets on the chemical composition of planet-hosting stars, we selected four wide binary systems in which at least one planet has been reported around one or both component stars.
For target selection, we cross-matched the wide binary catalog of \citet{El-Badry2021}, which is based on $Gaia$ astrometric solutions, with the March 2023 version of the exoplanet catalog from $\it{exoplanet.eu}$.
This cross-match yielded 180 wide binaries that host at least one confirmed planet, while the most recent catalog of planet-hosting binaries by \citet{Thebault2025} lists 759 systems.
Photometric magnitudes for each component star were obtained from 2MASS \citep{Skrutskie2006} and $Gaia$ Data Release 3 \citep[DR3;][]{GaiaCollaboration2023} to assess their observing feasibility.
We applied selection criteria requiring 2MASS $K$ $<$ 12.0 mag, $J-K$ $>$ 0.3, and $Gaia$ $BP-RP$ $>$ 0.65 for both components.
It should be noted that detailed NIR abundance analysis is not feasible for high-temperature stars due to the lack of absorption lines \citep{Lim2024}.
Considering sky positions and projected separations of the components, we initially selected several tens of planet-hosting wide binary candidates.

\begin{figure}
\centering
   \includegraphics[width=0.4\textwidth]{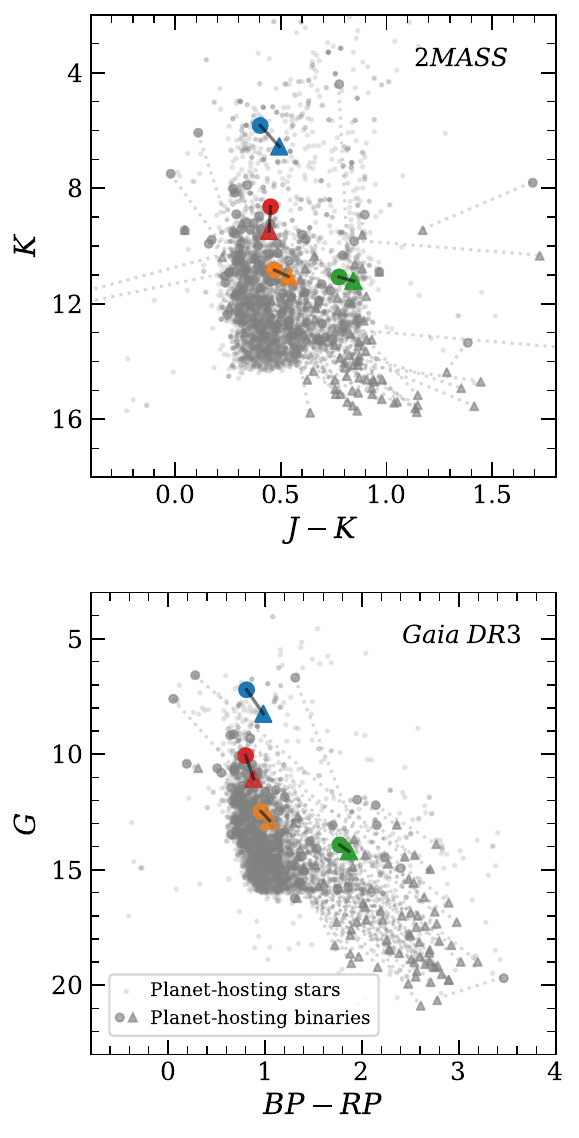}
     \caption{
     Our target wide binaries are marked with colored circles (primaries) and triangles (secondaries) in the 2MASS ($K$, $J-K$) and $Gaia$ ($G$, $BP-RP$) CMDs.
     Blue, orange, green, and red denote HD~20782/HD~20781, WASP-160~A/WASP-160~B, K2-54/K2-54~B, and WASP-127/TYC~4916-897-1 pairs, respectively, with the same color scheme adopted in subsequent figures.
     Grey circles and triangles indicate our initial candidate systems of planet-hosting wide binaries, while the smaller background symbols represent all reported planet-hosting stars to date.
     The two components of each binary are connected with solid and dotted lines.
     }
     \label{fig:cmd}
\end{figure}

We finally selected four wide binary systems with separations larger than 1000~au, in which both component stars exhibit similar colors and magnitudes in 2MASS ($K$, $J-K$) and $Gaia$ ($G$, $BP-RP$) color–magnitude diagrams (CMDs), as shown in Figure~\ref{fig:cmd}. The selected binaries are HD~20782/HD~20781, WASP-160~A/WASP-160~B, K2-54/Gaia~DR3~2594546351459477632 (hereafter K2-54~B\footnote{Since this star does not yet have a common identifier, we temporarily refer to it as K2-54~B.}), and WASP-127\footnote{Although BD$-$03~2978 is adopted as the primary identifier, WASP$-$127 is commonly used due to its planetary detection.}/TYC~4916-897-1, listed in the order of primary and secondary components.
We adopt a separation threshold of $>$1000~au to minimize potential dynamical influence of the companion on planet formation and evolution, as previous studies suggest that stellar multiplicity can affect planet occurrence up to separations of $\sim$100 $-$ 1500~au \citep{Wang2014, Hirsch2021}. 
In these CMDs, all known planet-hosting stars and our initial target wide binary candidates are also displayed for comparison. 
Most known planets are associated with the brighter primaries, while secondaries (typically fainter) tend to lie below the main locus of planet-hosting stars, implying that any planets around faint secondaries may remain undetected. 
Therefore, the similarity in colors and magnitudes between the two components suggests comparable chemical and dynamical evolution, while also ensuring that both stars are sufficiently bright for efficient planet searches.
Detailed information on our target stars is provided in Table~\ref{tab:target}. 
The two stars in each system are classified as G dwarfs (HD~20782/HD~20781 and WASP-127/TYC~4916-897-1), K dwarfs (WASP-160~A/WASP-160~B), and M dwarfs (K2-54/K2-54~B).

\begin{table*}
\small 
\setlength{\tabcolsep}{5pt}
\centering                                    
\caption{$Gaia$ DR3 astrometric parameters and comparison of $Gaia$ and our RV measurements for the target stars.}
\label{tab:target} 
\begin{tabular}{cccccccccc}   
\hline\hline        
\multirow{2}{*}{Name} 
& $\alpha$ (J2000) & $\delta$ (J2000) & Parallax  & $\mu_{\alpha}$  & $\mu_{\delta}$  &  RV ($Gaia$)  & RV (This study) & Sep \\
& [deg]            & [deg]            & [mas]     & [mas yr$^{-1}$] & [mas yr$^{-1}$] & [km~s$^{-1}$] & [km~s$^{-1}$]   & [au] \\
\hline
HD~20782       & 50.016679    & $-28.854363$  & 27.8760  & 349.054  & $-65.305$ & $39.85$ $\pm$ 0.12 & $40.35$ $\pm$ 1.67 & \multirow{2}{*}{   9086 }\\
HD~20781       & 50.014031    & $-28.784127$  & 27.8123  & 348.869  & $-66.614$ & $40.27$ $\pm$ 0.12 & $41.18$ $\pm$ 0.67 & \\
WASP-160~A     & 87.686408    & $-27.618245$  & 3.4608   & 26.870   & $-34.751$ & $-5.35$ $\pm$ 0.72 & $-5.52$ $\pm$ 1.62 & \multirow{2}{*}{   8245 } \\
WASP-160~B     & 87.679565    & $-27.623327$  & 3.4476   & 27.029   & $-34.802$ & $-6.26$ $\pm$ 0.94 & $-5.45$ $\pm$ 0.79 & \\
K2-54          & 338.054162   & $-17.544065$  & 5.7373   & $-5.018$ & $-9.804$  & $19.94$ $\pm$ 3.00 & $19.73$ $\pm$ 0.70 & \multirow{2}{*}{   1273 } \\
K2-54~B        & 338.054659   & $-17.546044$  & 5.7709   & $-5.979$ & $-9.493$  & $16.64$ $\pm$ 4.01 & $19.55$ $\pm$ 0.53 & \\
WASP-127       & 160.558767   & $-3.834996$   & 6.2241   & 19.133   & 17.058    & $-8.97$ $\pm$ 0.16 & $-9.17$ $\pm$ 1.32 & \multirow{2}{*}{   6509 } \\
TYC~4916-897-1 & 160.547657   & $-3.836883$   & 6.2140   & 18.769   & 16.489    & $-9.45$ $\pm$ 0.28 & $-8.12$ $\pm$ 1.70 & \\
\hline                                             
\end{tabular}
\end{table*}

\begin{table*}
\small 
\setlength{\tabcolsep}{3pt}
\centering
\caption{Planetary parameters for the planets orbiting our target stars.}
\label{tab:planet}
\begin{tabular}{cccccccccc}
\hline\hline
\multirow{2}{*}{Host} & \multirow{2}{*}{Planet} & \multirow{2}{*}{Type} & Discovery       & $P$      & $M_{\rm p}$                   & $R_{\rm p}$                    & $a$   & $e$ &\multirow{2}{*}{Reference} \\
                      &                         &                       & method          & [days]   & [$M_{\oplus}$ or $M_{\rm J}$] & [$R_{\oplus}$ or $R_{\rm J}$]  & [au]  &     & \\
\hline
HD~20782    & HD~20782~b    & Gas Giant     & Radial Velocity & 597.0643 & 1.4878 $M_{\rm J}$ & 1.21 $R_{\rm J}$  & 1.3649  & 0.950 & \citet{Udry2019} \\
HD~20781    & HD~20781~b    & Super-Earth   & Radial Velocity & 5.3135   & 1.93 $M_{\oplus}$  & 1.21 $R_{\oplus}$ & 0.0529  & 0.10  & \citet{Udry2019} \\
            & HD~20781~c    & Neptune-like  & Radial Velocity & 13.8905  & 5.33 $M_{\oplus}$  & 2.17 $R_{\oplus}$ & 0.1004  & 0.090 & \citet{Udry2019} \\
            & HD~20781~d    & Neptune-like  & Radial Velocity & 29.1580  & 10.61 $M_{\oplus}$ & 0.29 $R_{\rm J}$  & 0.1647  & 0.11  & \citet{Udry2019} \\
            & HD~20781~e    & Neptune-like  & Radial Velocity & 85.5073  & 14.03 $M_{\oplus}$ & 0.342 $R_{\rm J}$ & 0.3374  & 0.060 & \citet{Udry2019} \\
WASP-160~B  & WASP-160~B~b  & Gas Giant     & Primary Transit & 3.7685   & 0.278 $M_{\rm J}$  & 1.090 $R_{\rm J}$ & 0.0452  & 0.0   & \citet{Lendl2019} \\
K2-54       & K2-54~b       & Neptune-like  & Primary Transit & 9.7833   & 7.33 $M_{\oplus}$  & 2.61 $R_{\oplus}$ & 0.0761  & 0.24  & \citet{Thygesen2023} \\
WASP-127    & WASP-127~b    & Gas Giant     & Primary Transit & 4.1781   & 0.1647 $M_{\rm J}$ & 1.311 $R_{\rm J}$ & 0.0484  & 0.0   & \citet{Seidel2020} \\
\hline
\end{tabular}
\tablefoot{$P$, $M_{\rm p}$, $R_{\rm p}$, $a$, and $e$ denote the orbital period, planet mass, planet radius, semi-major axis, and orbital eccentricity, respectively. Planet masses and radii are expressed in Earth units for Super-Earth and Neptune-like planets, and in Jupiter units for gas giant planets. HD~20781~c lies near the transition between super-Earth and Neptune-like regimes based on its reported mass and radius.}
\end{table*}

The number of reported planets in our target systems ranges from one to four.
HD~20781 hosts four planets, while HD~20782 hosts one. 
The remaining three systems each contain a single known planet. 
Planetary types and parameters are summarized in Table~\ref{tab:planet}.
Although these four systems were selected primarily for their favorable observability and component similarity, the wide separations and diverse planetary environments make these systems well suited for testing whether planet formation and evolution leave detectable chemical imprints on their host stars.


\subsection{Observations} \label{sec:sub:obs}
Our spectroscopic observations were carried out from July to December 2023 under the support of the K-GMT Science Program (Program IDs: GS-2023B-Q-211 and GS-2023B-Q-315).
We employed the high-resolution NIR spectrograph IGRINS \citep{Mace2018}, mounted on the Gemini-South telescope.
IGRINS consists of two detectors that simultaneously cover the $H$ (1.49$-$1.80 $\mu$m) and $K$ (1.96$-$2.46 $\mu$m) bandpasses.
The average spectral resolving power is $R \sim 50,000$ in $H$-band and $R \sim 45,000$ in $K$-band.
We note that although IGRINS has since been moved to McDonald Observatory, its analogue instrument IGRINS-2 began operations at the Gemini-North telescope in late 2024 \citep{Oh2024}.
Both IGRINS and IGRINS-2 are widely used for exoplanet atmospheric studies via transmission spectroscopy \citep[e.g.,][]{Bazinet2024, Choi2025}.

Our observations were conducted under Band~2 and Band~3 weather conditions, with Band~2 corresponding to better transparency and image quality.
Depending on target brightness, WASP-160~A, WASP-160~B, K2-54, and K2-54~B were observed under Band~2 conditions, while HD~20782, HD~20781, and TYC~4916-897-1 were observed under Band~3.
Each target was observed using an ABBA or AB nodding sequence, with total exposure times ranging from 60 to 3700 seconds depending on stellar brightness.
In addition, a nearby A0V telluric standard star was observed immediately before or after each target to correct for atmospheric absorption features.

For WASP-127, we used archival data from the Raw and Reduced IGRINS Spectral Archive \citep[RRISA;][]{Sawczynec2025}, as this star had previously been observed to study the atmosphere of WASP-127~b under the Gemini program GS-2022A-LP-107.
The transmission spectroscopy observations were obtained in March 2022 and cover a full transit as well as an out-of-transit baseline \citep[see][]{Kanumalla2024}.
From this time-series dataset, we adopted the last spectroscopic data taken after the transit event.


\subsection{Data reduction} \label{sec:sub:reduction}
The basic reduction of the spectroscopic data was carried out using the IGRINS Pipeline Package version 3.0 \citep[PLP;][]{Kaplan2024}.
The PLP sequentially performs flat-fielding, bad-pixel correction, flexure correction, telluric correction, sky subtraction, and wavelength calibration, producing a one-dimensional spectrum from each nodding sequence.
The final spectrum for each star was obtained by merging the spectra over the effective wavelength ranges of the individual diffraction orders and applying continuum normalization.
Our data reduction procedures follow those described in \citet{Lim2025}.

Radial velocities (RVs) were measured through cross-correlation with a synthetic spectrum using the {\tt fxcor} task in the IRAF {\tt RV} package.
The typical uncertainty from the cross-correlation is about 1.1~km~s$^{-1}$.
The RVs listed in Table~\ref{tab:target} are barycentric-corrected values.
Our measurements agree well with those from $Gaia$ DR3, with a mean difference of 0.6~km~s$^{-1}$, except for K2-54~B.
Although this star shows the largest discrepancy (the $Gaia$ value is 2.9~km~s$^{-1}$ lower), the difference is consistent with the large $Gaia$ DR3 uncertainty of $\pm$4.0~km~s$^{-1}$ for this source.
These small RV variations measured at different epochs reduce the likelihood of an unseen tertiary companion in any of the systems, although the possibility of wider or lower-amplitude companions cannot be excluded.
The small RV differences between the components are 0.83, 0.07, 0.18, and 1.04~km~s$^{-1}$ for HD~20782/HD~20781, WASP-160~A/WASP-160~B, K2-54/K2-54~B, and WASP-127/TYC~4916-897-1 pairs, respectively, consistent with previous spectroscopic wide binary studies \citep[e.g.,][]{Quinn2009, Hawkins2020, Lim2024} and supporting that these pairs are genuine binaries despite their large separations (1000 to 9000~au).

The signal-to-noise ratios (SNRs) range from 110 to 220 per pixel and were estimated as the mean SNR measured at representative $H$-band (1.63~$\mu$m) and $K$-band (2.20~$\mu$m) regions.
For our targets, the SNR in the $K$-band spectra is generally higher than in the H band.


\section{Spectroscopic analysis} \label{sec:spec} 
\subsection{Atmospheric parameters} \label{sec:sub:atm}
Stellar atmospheric parameters, including effective temperature ($T_{\rm eff}$), surface gravity ($\log{g}$), microturbulence velocity ($v_{t}$), and [Fe/H], were derived using two independent methods.
The first is a canonical photometric method and the second is a grid-fitting technique based on synthetic spectra \citep[see][]{Lim2025}.
The resulting parameters, together with literature values for the planet-hosting stars in our sample, are summarized in Table~\ref{tab:param}.

For the photometric approach, we first estimated $T_{\rm eff}$ using the $Gaia$ $BP-RP$ color–temperature relation of \citet{Mucciarelli2021}, adopting reddening values from {\tt Bayestar19} \citep{Green2019}.
Surface gravity was then computed using
\begin{eqnarray} \label{eq:logg}
\log{g_{*}} = \log{g_{\odot}} + \log{\frac{M_{*}}{M_{\odot}}} 
+ 4\log{\frac{\rm T_{eff,*}}{\rm T_{eff,\odot}}} \nonumber 
+ 0.4(M_{bol,*} - M_{bol,\odot}), \nonumber
\end{eqnarray}
where bolometric corrections were computed using the {\tt Gaiadr3 Bcg} code \citep{Creevey2023} and stellar masses were estimated from \citet{Eker2018}.
Microturbulence was derived using the empirical relation of $v_{t}$ with $T_{\rm eff}$ and $\log{g}$ from \citet{Boeche2016}.
With these initial parameters, a first estimate of [Fe/H] was obtained through spectral synthesis based on MARCS model atmospheres \citep{Gustafsson2008}, and the procedure was iterated until the derived metallicity converged to the input value.

For the grid-fitting method, we employed a synthetic spectral grid (Yun et al., in preparation) generated using Kurucz’s NEWODF models based on the ATLAS9 atmosphere code \citep{Castelli2004}.
After matching the spectral resolution and normalizing both synthetic and observed spectra, we identified the best-fitting model by minimizing residuals between the two.
Because the synthetic grid spans a wide range in $T_{\rm eff}$ (3,500$-$8,000~K with 250~K steps), $\log{g}$ (0.0$-$5.0~dex with 0.25~dex steps), and [Fe/H] ($-3.0$ to $+1.0$~dex with 0.25~dex steps), these three parameters could be determined simultaneously.
The value of $v_{t}$ was subsequently obtained using the \citet{Boeche2016} relation.
To ensure consistency with our abundance analysis, [Fe/H] was rederived through spectral synthesis of individual Fe lines using the determined atmospheric parameters, following the same procedure as in the photometric method.
The re-estimated [Fe/H] values are typically about 0.15~dex higher than those directly obtained from the grid-fitting technique, likely reflecting differences in line selection, fitting strategy, and adopted model atmospheres.

\begin{figure}
\centering
   \includegraphics[width=0.48\textwidth]{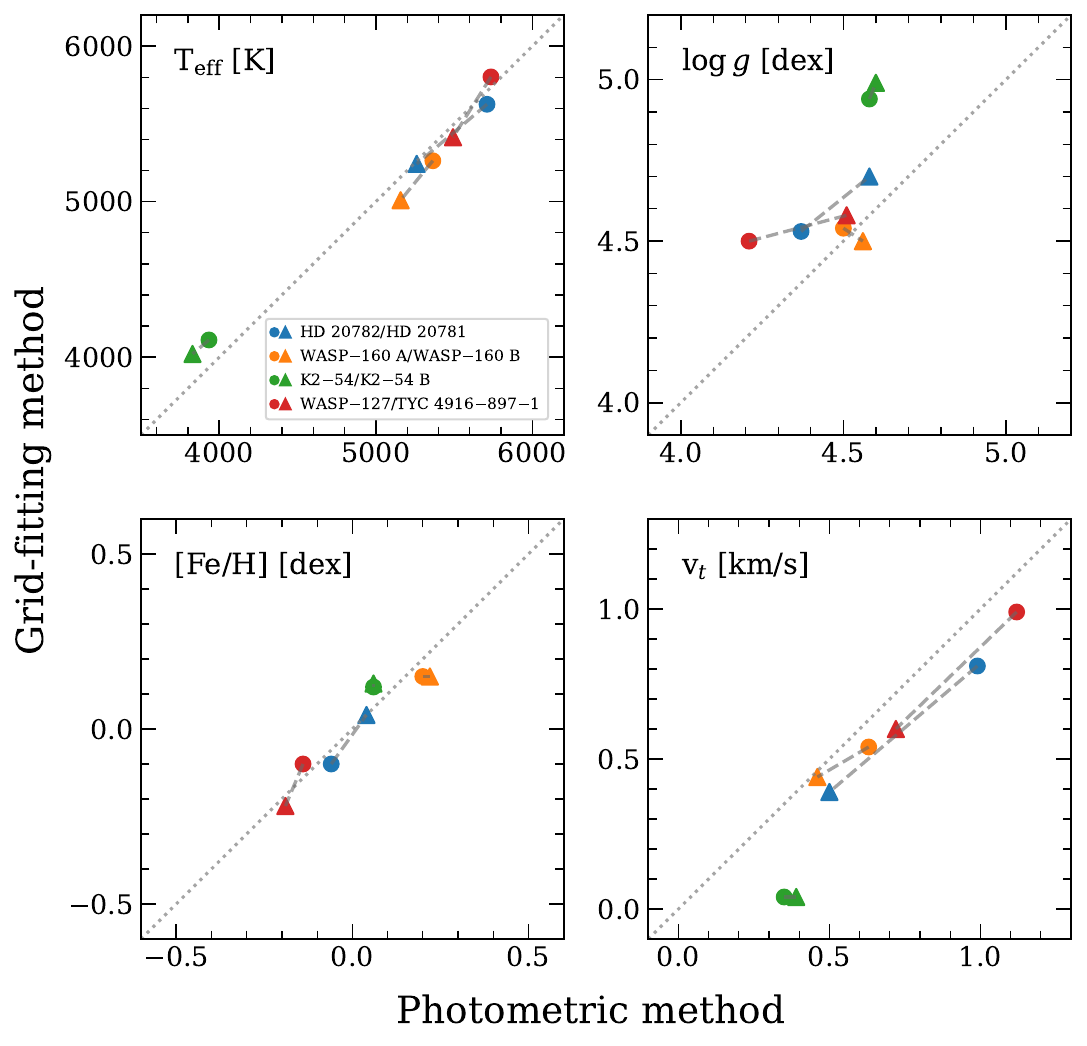}
     \caption{
     Comparison of atmospheric parameters derived from the photometric and grid-fitting methods.
     Each wide binary pair is identified by a different color (see legend in the upper-left panel).
     The dotted line indicates the one-to-one relation.
     While $\log{g}$ and $v_{t}$ exhibit some offsets between the two methods, $T_{\rm eff}$ and [Fe/H] show good overall agreement.
     The small differences in [Fe/H] arise because $T_{\rm eff}$ is the primary parameter that influences chemical abundance measurements.
     }
     \label{fig:param}
\end{figure}

Figure~\ref{fig:param} and Table~\ref{tab:param} compare atmospheric parameters derived from the two independent approaches.
Overall, the two methods show good agreement in $T_{\rm eff}$ and [Fe/H], while systematic offsets appear in $\log{g}$ and $v_{t}$.
In particular, $\log{g}$ values obtained from the grid-fitting method are generally larger, whereas $v_{t}$ values are systematically smaller than those derived from the photometric method.
The median differences are 92~K in $T_{\rm eff}$, 0.14~dex in $\log{g}$, 0.12~km~s$^{-1}$ in $v_{t}$, and 0.045~dex in [Fe/H].
The discrepancies are most pronounced for the K2-54/K2-54~B pair, which has the lowest temperatures in our sample, whereas the remaining stars show close agreement between the two estimates.
This overall agreement supports the robustness of our atmospheric parameter determinations, despite the fact that the two methods rely on different underlying model atmospheres, namely MARCS and Kurucz.
Although chemical abundances were computed using both parameter sets, we adopted the photometric parameters as our primary atmospheric parameters to maintain consistency with the abundance analysis.
This choice is motivated by consistency with our abundance determination scheme and by the closer agreement of the photometric parameters with literature values, particularly in $\log{g}$ (see Table~\ref{tab:param}).
We also note that the abundance differences between binary components are slightly smaller when adopting the photometric parameters than when using the grid-fitting results (see Section~\ref{sec:sub:abund}).

\begin{table*}
\small 
\setlength{\tabcolsep}{3pt}
\centering
\caption{Atmospheric parameters derived from the photometric and grid-fitting methods, and literature values.}
\label{tab:param}
\begin{tabular}{@{\extracolsep{4pt}}c cccc cccc cccc}
\hline\hline
\multirow{3}{*}{Name} & \multicolumn{4}{c}{Photometric} & \multicolumn{4}{c}{Grid-fitting} & \multicolumn{4}{c}{Literature} \\
\cline{2-5} \cline{6-9} \cline{10-13}
                      & $T_{\rm eff}$ & $\log{g}$ & $v_{t}$ & [Fe/H] &
                        $T_{\rm eff}$ & $\log{g}$ & $v_{t}$ & [Fe/H] &
                        $T_{\rm eff}$ & $\log{g}$ & [Fe/H] & Ref. \\
                      & [K] & [dex] & [km~s$^{-1}$] & [dex] &
                        [K] & [dex] & [km~s$^{-1}$] & [dex] &
                        [K] & [dex] & [dex] & \\
\hline
HD~20782       & 5710 & 4.37 & 0.99 & $-0.06$ & 5626 & 4.53 & 0.81 & $-0.10$ & 5790 & 4.38 & $-0.11$ & \citet{Stassun2017} \\
HD~20781       & 5261 & 4.58 & 0.50 & $0.04$  & 5243 & 4.70 & 0.39 & $0.04$  & 5256 & 4.37 & $-0.11$ & {\footnotesize\citet{Udry2019}} \\
WASP-160~A     & 5364 & 4.50 & 0.63 & $0.20$  & 5263 & 4.54 & 0.54 & $0.15$  & 5424 & 4.42 & $0.14$  & \citet{Jofre2021} \\
WASP-160~B     & 5158 & 4.56 & 0.46 & $0.22$  & 5009 & 4.50 & 0.44 & $0.15$  & 5215 & 4.47 & $0.15$  & \citet{Jofre2021} \\
K2-54          & 3934 & 4.58 & 0.35 & $0.06$  & 4111 & 4.94 & 0.04 & $0.12$  & 3990 & 4.61 & $-0.02$ & {\small \citet{Thygesen2023}} \\
K2-54~B        & 3829 & 4.60 & 0.39 & $0.06$  & 4021 & 4.99 & 0.04 & $0.13$  & --   & --   & --      & -- \\
WASP-127       & 5735 & 4.21 & 1.12 & $-0.14$ & 5802 & 4.50 & 0.99 & $-0.10$ & 5620 & 4.18 & $-0.18$ & \citet{Lam2017} \\
TYC~4916-897-1 & 5493 & 4.51 & 0.72 & $-0.19$ & 5414 & 4.58 & 0.60 & $-0.22$ & --   & --   & --      & -- \\
\hline
\end{tabular}
\end{table*}

On the other hand, the two component stars in each wide binary pair show very similar values in all atmospheric parameters, consistent with their nearly coincident positions in the CMDs shown in Figure~\ref{fig:cmd}.
The K2-54/K2-54~B pair exhibits the smallest differences, with $\Delta T_{\rm eff}$ = 105/90~K, $\Delta \log{g}$ = 0.02/0.05~dex, and $\Delta v_{t}$ = 0.04/0.00~km~s$^{-1}$ (photometric/grid-fitting values).
Even though the HD~20782/HD~20781 pair shows somewhat larger offsets, the differences remain modest ($\Delta T_{\rm eff} \sim$ 400~K, $\Delta \log{g} \sim$ 0.2~dex, and $\Delta v_{t} \sim$ 0.45~km~s$^{-1}$).
The close similarity in atmospheric parameters between the components, within their comparable stellar masses, supports the interpretation that these stars form genuine binaries that originated from the same birth environment and share similar evolutionary stages.


\subsection{Chemical abundance measurement} \label{sec:sub:chem}
Based on the reduced one-dimensional spectra for each star, we examined detailed chemical abundances for 18 elements (C, N, O, Na, Mg, Al, Si, P, S, K, Ca, Sc, Ti, Cr, Fe, Co, Ni, and Yb) that are measurable within our NIR spectral range.
As a first step, we generated model atmospheres for each star using plane-parallel MARCS models \citep{Gustafsson2008} with the atmospheric parameters derived in Section~\ref{sec:sub:atm}.
Chemical abundances were measured via spectral synthesis of individual lines using the 2019NOV version of the local thermodynamic equilibrium (LTE) code MOOG \citep{Sneden1973}.
The atomic and molecular line lists were adopted from \citet{Lim2022, Lim2025}.

As described in Section~\ref{sec:sub:atm}, Fe abundances were determined first from \ions{Fe}{i} lines during the process of atmospheric parameter determination.
Following the procedure of \citet{Lim2025}, the abundances of C, N, and O were then derived from CO, CN, and OH molecular features, respectively.
We performed three iterative measurements in the order of  O, C, and N, updating the input elemental abundances in the model atmosphere after each step.
For C, N, and O, the NIR spectra provide numerous molecular features, typically more than 10 usable lines for each element, which offers a practical advantage over the optical, where these elements are often measured from only a few lines ($\lesssim$ 3) or remain less well constrained.
The abundances of all other elements were then derived through iterative spectral synthesis with corresponding updates to the model atmosphere.
Final abundances were computed as the mean values from multiple absorption lines, and the uncertainties were adopted as the standard error of the mean.
The abundance ratios derived using the photometric atmospheric parameters, adopted as our final values, as well as those obtained with the grid-fitting parameters, are listed in the [X/H] scale relative to the solar abundances of \citet{Asplund2009} in Tables~\ref{tab:abund_HD20782}--\ref{tab:abund_WASP127}.

Unlike many optical studies that employ strictly differential techniques \citep[e.g.,][]{Desidera2004, Saffe2016, Maia2019, Flores2024}, our analysis is based on consistent line-by-line spectral synthesis of common lines within each wide binary system, enabling robust relative abundance comparisons between the two components.
This approach is motivated by the use of photometrically determined atmospheric parameters and by the more prominent molecular features in the NIR spectral range.

In order to evaluate the effect of non-LTE corrections on the chemical abundances of our target stars, we estimated line-by-line non-LTE abundance corrections\footnote{\url{https://nlte.mpia.de}} for Fe \citep{Bergemann2012} based on the individual atmospheric parameters of each star.
The non-LTE corrections introduce only a mean difference of 0.005~dex relative to the LTE values, with non-LTE abundances being slightly higher.
Because this difference is negligible compared to the measurement uncertainties, and has only a minor influence on our primary goal of comparing abundances between the binary components, we adopt the LTE abundances in the following analysis.

For five of our target stars (HD~20782/HD~20781, WASP-160~A/WASP-160~B, and WASP-127), literature abundances compiled in the {\it Hypatia Catalog} \citep{Hinkel2014} allow direct comparison.
Our [Fe/H] measurements differ from literature values by typically 0.06~dex, with a maximum deviation of 0.11~dex for HD~20781.
Across all elements in common, the mean absolute difference is 0.13~dex, although individual cases show deviations up to $\sim$0.5~dex.
The largest discrepancies are associated with elements measured from only a few spectral lines, such as Sc and Co, while most elements exhibit differences below 0.15~dex.
Notably, N and O abundances show comparatively large deviations ($>$ 0.3~dex) for WASP-160~A and WASP-127.
Such differences are expected because the referenced studies employed different instruments, methods, wavelength coverage, adopted temperature scale, and adopted model atmospheres, $-$ in contrast to our abundances, which are derived exclusively from NIR spectra $-$ and therefore systematic offsets may arise.
Given these limitations and our focus on relative abundance differences between components within the same binary systems, we adopt the abundances derived in this work for the main analysis, and use literature values to compare abundance differences between the binary components rather than their absolute abundance scales.


\section{Chemical properties of wide binary systems} \label{sec:result} 
\subsection{Chemical differences between components} \label{sec:sub:abund}

As discussed in Section~\ref{sec:intro}, wide binary component stars are expected to be chemically homogeneous if they were born together from the same molecular cloud and have evolved without significant external perturbations. 
Therefore, any measurable abundance difference between the components can provide important clues to the peculiar chemical or dynamical history of the system, including possible signatures related to planet formation or evolutionary processes.

To investigate these effects, we examined the abundance differences between the two stars in each binary, determined as the abundance of the primary minus that of the secondary. 
The differences in [Fe/H] are $-0.104$, $-0.021$, $-0.003$, and $0.049$~dex for the HD~20782/HD~20781, WASP-160~A/WASP-160~B, K2-54/K2-54~B, and WASP-127/TYC~4916-897-1 systems, respectively.
These differences correspond to 4.1$\sigma$, 0.6$\sigma$, 0.1$\sigma$, and 1.2$\sigma$ relative to the measurement uncertainties for each system.
Three systems show small differences ($|\Delta$[Fe/H]$|$ $<$ 0.05~dex within $<$1.5$\sigma$), consistent with previous high-resolution studies of chemically homogeneous wide binaries \citep[e.g.,][]{Nelson2021}.
In contrast, HD~20782/HD~20781 pair displays a noticeably larger Fe abundance difference of 0.10~dex.
We note that this system has the largest binary separation ($\sim$9000~au) and the highest planetary multiplicity among our sample (four planets orbiting the secondary and one orbiting the primary; see Table~\ref{tab:planet}).

\begin{figure}
\centering
   \includegraphics[width=0.49\textwidth]{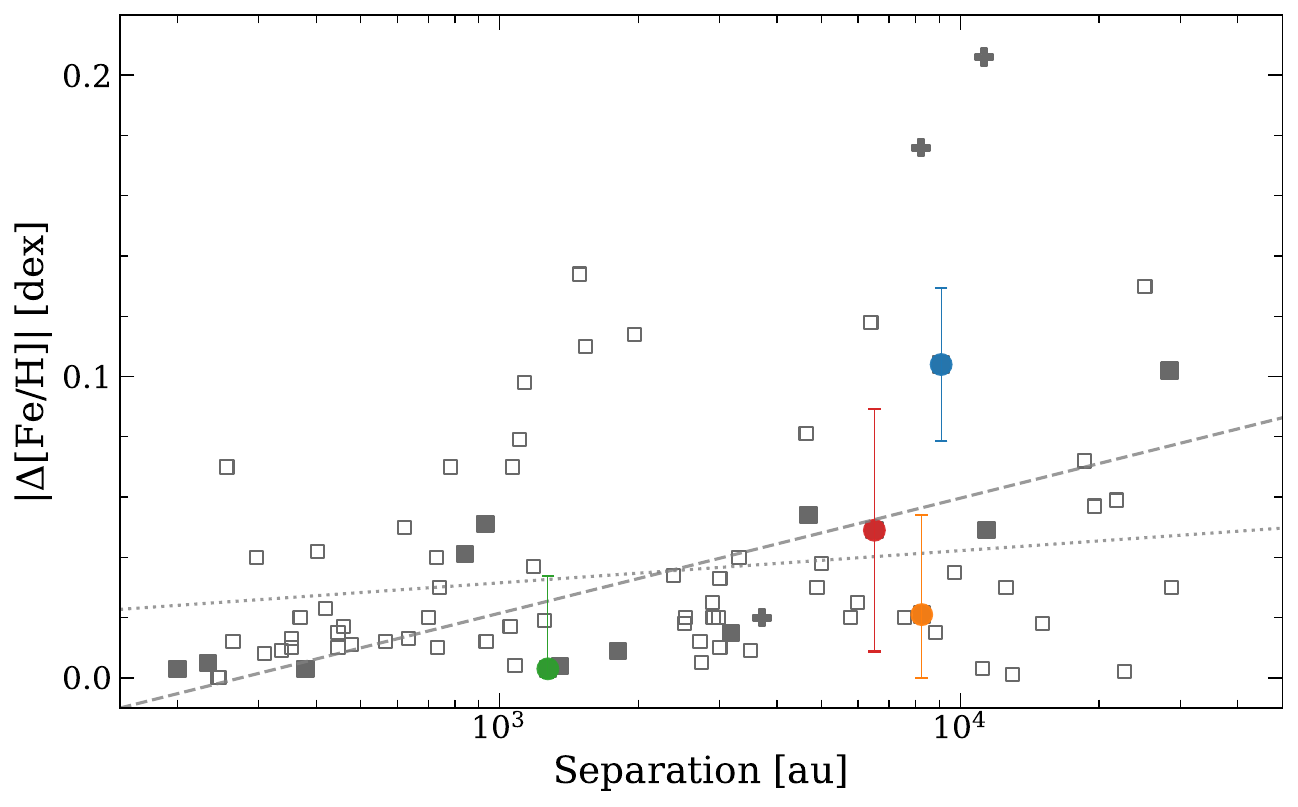}
     \caption{
     Absolute differences in [Fe/H] as a function of binary separation. 
     Our sample is shown with colored symbols, while comparison data from the literature \citep{Ramirez2019, Hawkins2020, Lim2021, Lim2024, Nagar2020, Nelson2021, Liu2021} are plotted as square symbols, where filled and open squares represent binaries with and without reported planets, respectively.
     The three plus symbols denote wide binaries with indirect evidence for planetary signatures, despite the absence of confirmed planets.
     For systems included in both our study and the literature, our measured values are used.
     The dashed and dotted lines show least-squares fits to the samples with and without currently confirmed planets, with the fit performed as a linear function of the logarithm of the projected separation. 
     Although the fit for binaries with confirmed planets appears steeper, this may partly reflect small-number statistics.
     Accordingly, we defer any interpretation of the fitted slopes until a larger and more representative sample becomes available.
     }
     \label{fig:sep_feh}
\end{figure}

Previous work has suggested a possible trend of increasing $\Delta$[Fe/H] with wider separations \citep[e.g.,][]{Ramirez2019}, although the scatter also becomes larger at wide separations \citep{Lim2024}. 
As shown in Figure~\ref{fig:sep_feh}, the Fe difference of HD~20782/HD~20781 pair is still relatively high compared to other binaries with similar separations, regardless of whether they host planets. 
Extreme cases exist in the literature, such as the HD~240430/HD~240429 (Kronos/Krios) pair, which shows a $\ge$ 0.2~dex difference at a separation of $\sim$10,000~au that has been interpreted as a signature of planet engulfment \citep{Oh2018}. 
However, the Fe abundance alone is insufficient to attribute the difference in HD~20782/HD~20781 solely to planet-related processes, and a more comprehensive interpretation requires examination of a broader set of elements.
Finally, we note that literature values for $\Delta$[Fe/H] reported by \citet{Mack2014} for HD~20782/HD~20781 ($-0.06$~dex) and by \citet{Lendl2019} and \citet{Jofre2021} for WASP-160~A/WASP-160~B ($-0.08$~dex and $-0.01$~dex, respectively) differ slightly from our estimates.

\begin{figure}
\centering
   \includegraphics[width=0.48\textwidth]{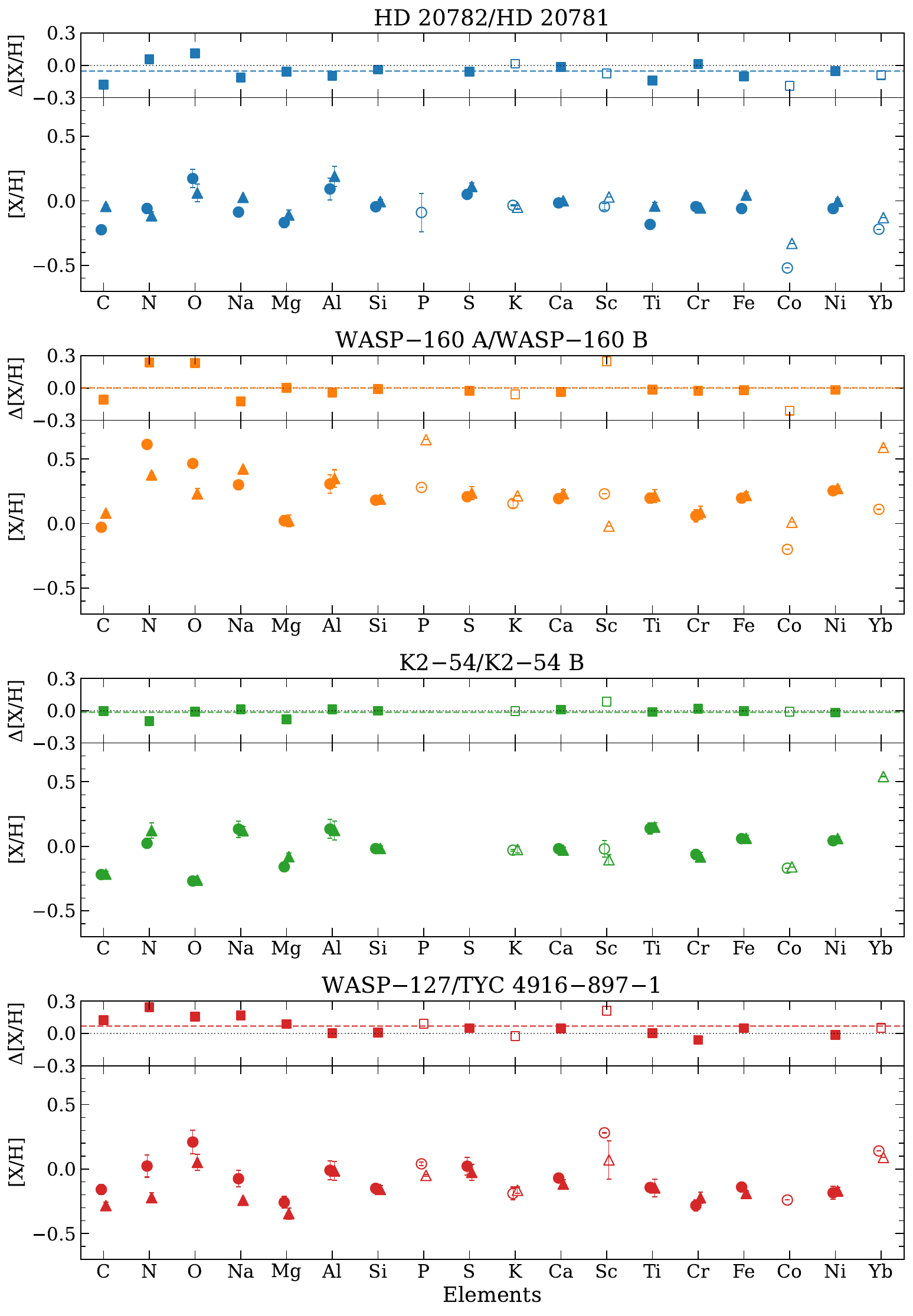}
     \caption{
     Elemental abundance ratios [X/H] for the components of each wide binary system.
     The lower plots of each panel show the absolute abundances of the primary and secondary stars, displayed as circles and triangles, respectively, with vertical bars indicating $\pm 1~\sigma$ measurement uncertainties.
     The upper plots show the abundance differences $\Delta$[X/H].
     Filled symbols denote the 13 reliable elements used in the quantitative comparison (see main text).
     In the upper plots, the colored dashed line marks the mean $\Delta$[X/H] derived from the reliable element set, and the dotted line indicates zero difference.
     }
     \label{fig:delta_XH}
\end{figure}

To investigate chemical differences on an element-by-element basis, we compared the [X/H] abundance ratios of the two stars in each binary system and examined the corresponding differential abundances, defined as $\Delta$[X/H] = [X/H]$_{\rm primary}$ $-$ [X/H]$_{\rm secondary}$, as presented in Table~\ref{tab:delta_abund} and shown in Figure~\ref{fig:delta_XH}.
At first glance, the K2-54/K2-54~B system exhibits minimal differences across all elements, whereas the remaining three pairs show noticeable differences for certain species.
For a more quantitative evaluation, we selected 13 reliable elements (C, N, O, Na, Mg, Al, Si, S, Ca, Ti, Cr, Fe, and Ni), excluding elements measured from fewer than three lines in order to reduce the influence of less secure measurements (see Appendix~\ref{app:robust} and Tables~\ref{tab:abund_HD20782}--\ref{tab:abund_WASP127}). 
The average differential abundances among the reliable elements, $\overline{\Delta\mathrm{[X/H]}}$, are $-0.053$, $0.003$, $-0.014$, and $0.067$~dex for HD~20782/HD~20781 through WASP-127/TYC~4916-897-1, respectively, while the mean absolute differences, $\overline{|\Delta\mathrm{[X/H]}|}$, are 0.081, 0.069, 0.023, and 0.078~dex. 
To assess the significance of these differences, we estimated the average measurement uncertainties for the same element set, which are typically $\sim$0.03~dex for six stars, and slightly larger for the WASP~127 and TYC~4916-897-1 (0.05 and 0.04~dex).
As suggested by Figure~\ref{fig:delta_XH}, K2-54/K2-54~B shows the smallest level of variation, consistent with strong chemical homogeneity. 
For the HD~20782/HD~20781 and WASP-127/TYC~4916-897-1 systems, the mean differential abundance and the mean absolute difference are very similar, implying that the abundance offsets are systematic rather than random. 
In HD~20782/HD~20781, the secondary (HD~20781) is systematically more enriched, whereas in WASP-127/TYC~4916-897-1, the primary (WASP-127) shows modest enhancement.
In contrast, WASP-160~A/WASP-160~B system exhibits a small mean differential abundance ($\overline{\Delta\mathrm{[X/H]}} = 0.003$~dex) but a large mean absolute difference ($\overline{|\Delta\mathrm{[X/H]}|} = 0.069$~dex), indicating element-by-element variation with opposite signs. 
This behavior is consistent with the findings of \citet{Jofre2021}, who reported opposite abundance trends between volatile and refractory element groups (see Section~\ref{sec:sub:WASP-160} for further details).

Overall, the level of chemical differences in our sample (0.02$\sim$0.08~dex) is comparable to those reported in previous studies of wide binaries without reported planets \citep[e.g.,][]{Hawkins2020, Lim2024}, suggesting that the presence of planets alone does not necessarily lead to significant chemical divergence between binary components. 
These results support the view that wide binaries are chemically homogeneous, consistent with a common formation origin, regardless of whether they host planets.

On the other hand, as described in Section~\ref{sec:sub:atm}, we determined atmospheric parameters using two independent approaches (photometric and grid-fitting).
When chemical abundances were derived using the grid-fitting parameter set, the abundance differences between binary components became moderately larger, although the overall trends remained consistent with the photometric-based results.
Using the grid-fitting parameters, the differences in [Fe/H] are $-0.139$, $-0.006$, $-0.007$, and $0.121$~dex, and the mean absolute differences among the 13 reliable elements are 0.125, 0.062, 0.022, and 0.144~dex for the HD~20782/HD~20781, WASP-160~A/WASP-160~B, K2-54/K2-54~B, and WASP-127/TYC~4916-897-1 systems, respectively.
These values are larger than those obtained with the photometric parameters for HD~20782/HD~20781 and WASP-127/TYC~4916-897-1, while smaller or comparable for WASP-160~A/WASP-160~B and K2-54/K2-54~B.
The dependence of the abundance differences on the adopted atmospheric parameters may represent a limitation on the robustness of the analysis, not only in this study but also in many spectroscopic studies.
Nevertheless, the consistently small differences and the preserved abundance patterns suggest that homogeneous and internally consistent analysis provides a stable basis for differential comparisons between binary components.


\subsection{Trends of chemical differences with elemental condensation temperature} \label{sec:sub:Tcond}

To examine element-by-element chemical differences within each wide binary system, we arrange the elements in order of their condensation temperature ($T_{\rm cond}$), adopting the values from \citet[][see also \citealt{Lodders2003}]{Lodders2025}.
Following these references, the elements are divided into three groups according to $T_{\rm cond}$: volatile ($T_{\rm cond} < 660~\mathrm{K}$), moderately volatile ($660~\mathrm{K} < T_{\rm cond} < 1300~\mathrm{K}$), and refractory ($T_{\rm cond} > 1300~\mathrm{K}$).
Differences between the volatile and refractory element groups, or more generally trends of the abundance differences with $T_{\rm cond}$ between the two components, are commonly interpreted as signatures of planet-star interactions \citep[e.g.][]{Oh2018, Jofre2021}.

\begin{figure}
\centering
   \includegraphics[width=0.48\textwidth]{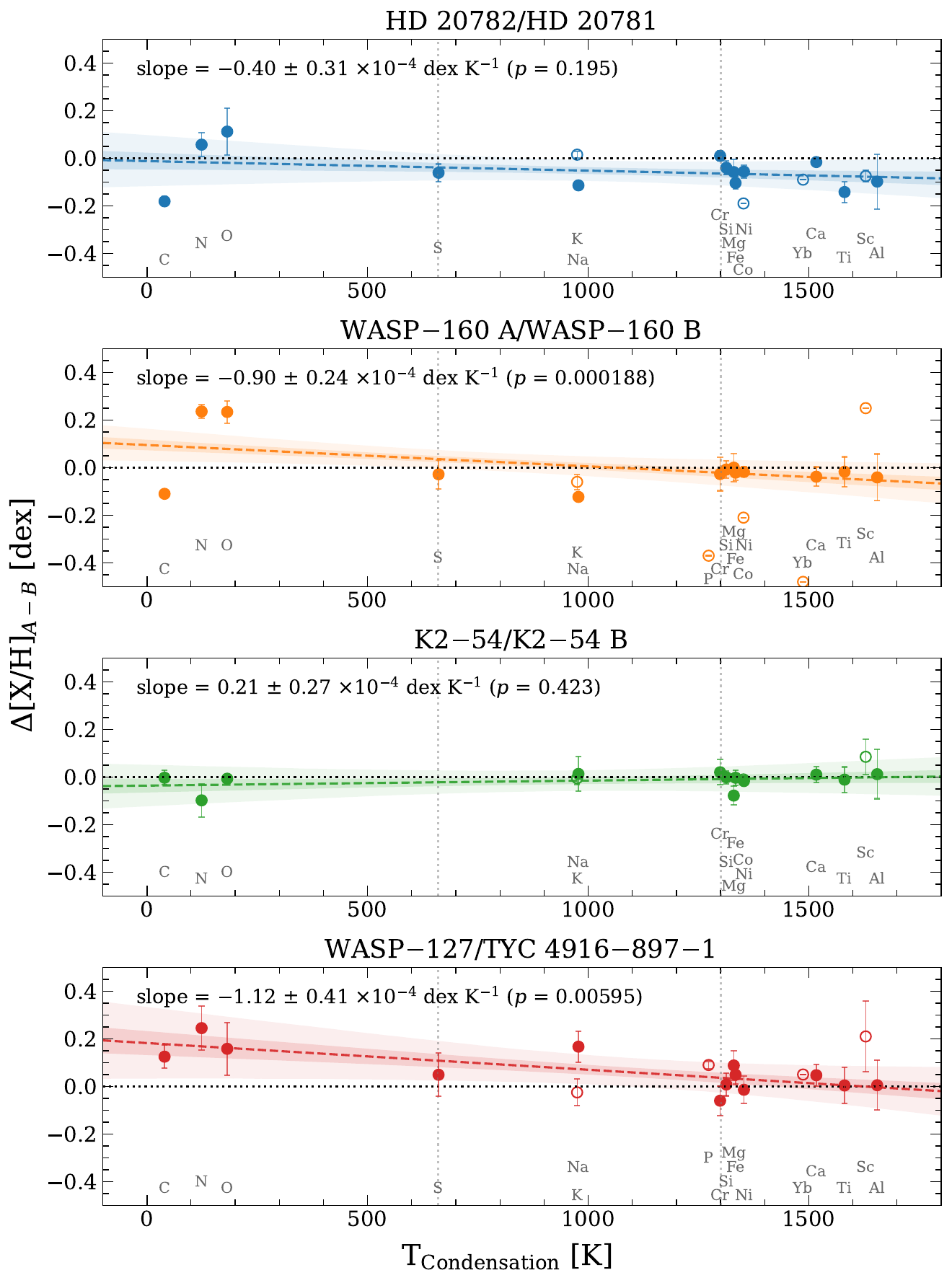}
     \caption{
     Abundance differences between the two components of the four planet-hosting wide binaries as a function of $T_{\rm cond}$.
     The vertical dotted lines mark $T_{\rm cond} = 660$~K and 1300~K, which separate volatile, moderately volatile, and refractory elements.  
     The corresponding elements are indicated at the bottom of each panel.
     Filled circles indicate elements with reliable abundance measurements that are used in the fit, while open circles mark elements that are excluded from the regression.
     Error bars correspond to the 1$\sigma$ uncertainties of $\Delta$[X/H].
     The colored dashed line in each panel shows the median linear relation obtained from the Monte Carlo regression, and the shaded regions indicate the 1$\sigma$ (dark) and 3$\sigma$ (light) confidence bands of the fitted relation.
     The best fitting slopes and the two-sided $p$-values for the null hypothesis of zero slope are given in the upper left corner of each panel.
     The small $p$-values measured for the WASP-160~A/WASP-160~B and WASP-127/TYC~4916-897-1 systems indicate statistically significant correlations, whereas the large $p$-value obtained for K2-54/K2-54~B is consistent with a null slope.
     }
     \label{fig:T_cond}
\end{figure}

Figure~\ref{fig:T_cond} shows the trend of the abundance differences between the two components as a function of $T_{\rm cond}$ for our four planet-hosting wide binary systems.
As in the previous subsection, the differences are defined as primary minus secondary values, while the uncertainties are estimated as the quadratic sum of the errors of the two stars for each element.
In order to quantify the trend with $T_{\rm cond}$ and evaluate the significance of any correlation, we perform a Monte Carlo analysis (N = 50,000) that takes into account the measurement uncertainties in the abundance differences.
In each realization, abundances are perturbed within their errors and a linear relation is fitted to the 13 reliable elements, as our primary goal is to quantify the overall slope of the differential abundances as a function of $T_{\rm cond}$.
We adopt the median and the 1$\sigma$ scatter of the resulting slope distribution as the best-fitting slope and its uncertainty. 
Approximating the slope distribution as normal, 
we further assess the statistical significance of the slope by computing a two-sided $p$-value for the null hypothesis of zero slope from the Monte Carlo realizations, where values closer to zero indicate a more significant correlation between the abundance differences and $T_{\rm cond}$.

Among our four systems, the WASP-160~A/WASP-160~B and WASP-127/TYC~4916-897-1 pairs show clear and statistically significant trends with $T_{\rm cond}$, while the trend for HD~20782/HD~20781 is only marginal and K2-54/K2-54~B is consistent with a flat relation.
The corresponding slopes and $p$-values are listed in each panel of Figure~\ref{fig:T_cond}.
It is important to note that the trends with $T_{\rm cond}$ differ among the four systems, even though each system contains at least one planet-hosting star.
This diversity may suggest that additional factors contribute to the observed trends, or that any planetary signature may vary from one system to another.

For both WASP-160~A/WASP-160~B and WASP-127/TYC~4916-897-1, the inferred slopes are negative and significant at about the 3.8$\sigma$ and 2.7$\sigma$ levels, respectively, according to our Monte Carlo analysis.
A closer inspection of Figure~\ref{fig:T_cond} reveals that, in these two systems, the abundance differences are larger for volatile elements such as C, N, and O, while those for refractory elements are close to zero.
In other words, the secondary stars appear more depleted in volatile elements, but show nearly identical refractory abundances compared to their primaries.
It is noteworthy that a gas giant planet (WASP-160~B~b) orbits the secondary star in the WASP-160~A/WASP-160~B pair, whereas the primary hosts a gas giant (WASP-127~b) in the WASP-127/TYC~4916-897-1 system, and no planet has been reported around the remaining components.
If these significant trends with $T_{\rm cond}$ in both systems are related to star-planet interactions, the opposite host-star configurations may suggest that the underlying mechanisms, and their effects on the host stellar atmospheres, differ between the two systems.
A more detailed discussion of each system is presented in the following section.
In contrast, the K2-54/K2-54~B system shows no significant chemical trend with $T_{\rm cond}$, despite the presence of a Neptune-like planet around the primary star.
The HD~20782/HD~20781 pair, hosting one and four planets, respectively, similarly shows at most a weak correlation.
Taken together, these results indicate that the chemical signatures imprinted on planet-host stars by their planetary systems can be diverse, and that planetary effects may not be the only driver of the differential abundances between the components, highlighting the need for detailed system-by-system investigations of individual planet-host stars in wide binary systems.

We also examined how the inferred $T_{\rm cond}$ slopes change when the grid-fitting atmospheric parameters are adopted and when all measured elements are included in the fit.
In both cases, the overall trends remain qualitatively consistent with those obtained from the primary analysis based on the photometric parameter set and the 13 reliable elements.
The detailed comparison is presented in Appendix~\ref{app:robust}.


\section{Chemical signatures in individual systems} \label{sec:result2}

As the four systems show diverse differential abundance trends with $T_{\rm cond}$, their physical interpretation may not be uniform from one system to another.
In this section, using the differential abundance trends with $T_{\rm cond}$, we discuss the origin of the observed chemical patterns on a system-by-system basis, considering both planet-related scenarios and alternative stellar processes that may influence the host stellar atmospheres.

\begin{figure}
\centering
   \includegraphics[width=0.48\textwidth]{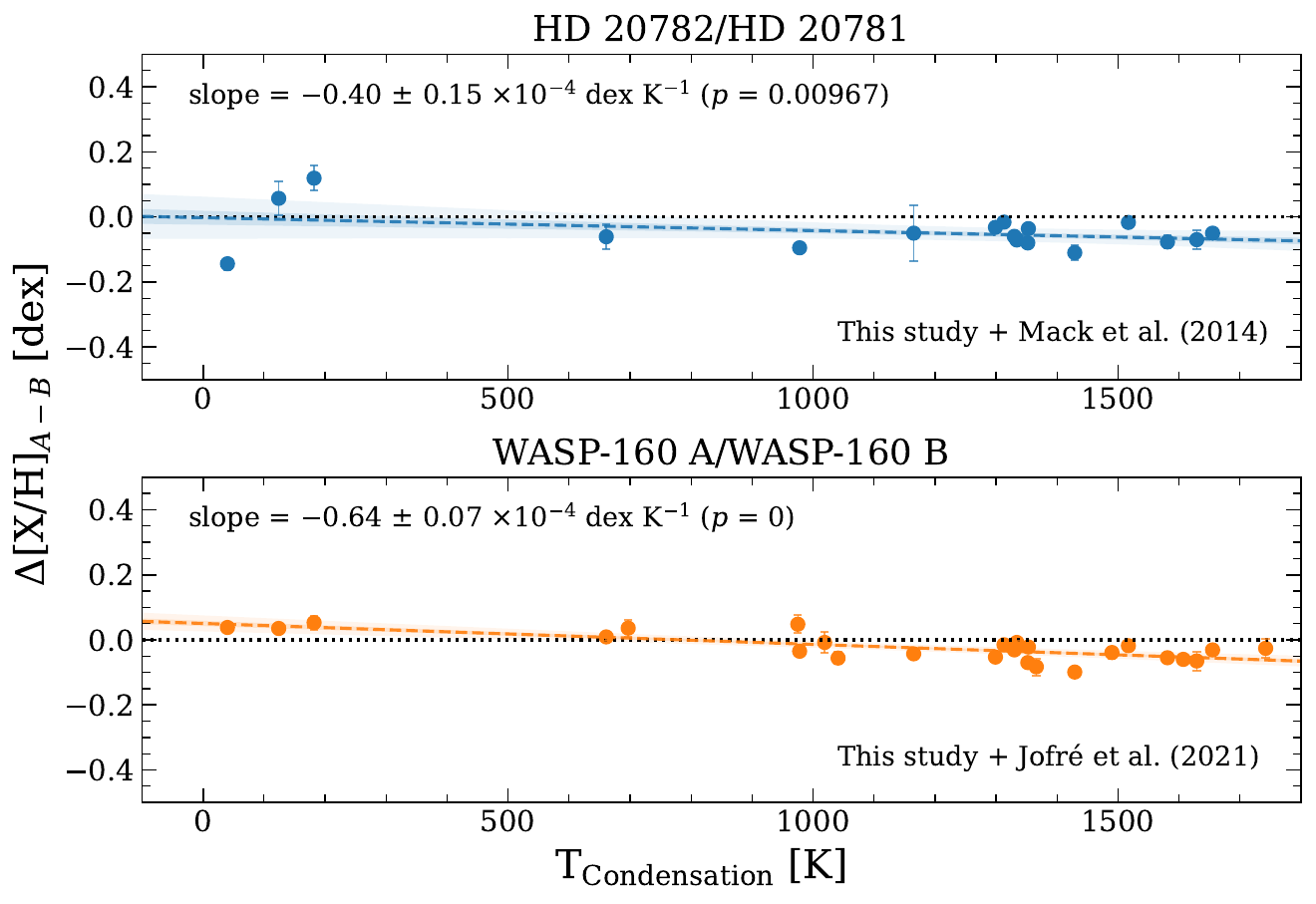}
     \caption{
     Same as Figure~\ref{fig:T_cond}, but including literature measurements combined with our data using error-weighted means.
     For the HD~20782/HD~20781 and WASP-160~A/WASP-160~B systems, data from \citet{Mack2014} and \citet{Jofre2021} are included, respectively.
     While the overall shapes and signs of the $T_{\rm cond}$ trends remain consistent with those derived from our data alone, the statistical significance of the correlations is substantially enhanced by the inclusion of the literature measurements.
     }
     \label{fig:T_cond_combine}
\end{figure}

\subsection{HD~20782/HD~20781} \label{sec:sub:HD20782}

The HD~20782/HD~20781 system is a rare example of a wide binary in which both components host planetary systems, yet with markedly different architectures.
HD~20782 hosts a single gas giant on an extremely eccentric orbit ($e = 0.950$), whereas HD~20781 harbors a super-Earth and three Neptunian planets (see Table~\ref{tab:planet}).
This configuration provides a valuable laboratory to investigate how different planetary environments may affect chemical signatures in host stellar atmospheres.

Previous high-resolution spectroscopy by \citet{Mack2014} reported enhanced refractory abundances in both HD~20782 and HD~20781 stars, as well as positive correlations between [X/H] and $T_{\rm cond}$ for elements with $T_{\rm cond} > 900$~K.
These trends were interpreted as evidence for the ingestion of 10$-$20~M$_\oplus$ of rocky material by each star\footnote{The possible impact of the two Neptune-like planets around HD~20781 was not discussed in \citet{Mack2014}, as these planets were discovered later through RV monitoring by \citet{Udry2019}.}.
Although our data also show similar positive [X/H]-$T_{\rm cond}$ correlations for each star individually, we focus on the differential abundance patterns between the two components, since absolute abundances of individual stars can be affected by Galactic chemical evolution and other environmental effects \citep{Adibekyan2014, Onehag2014}.

In the upper panel of Figure~\ref{fig:T_cond_combine}, we re-examine the abundance differences $\Delta$[X/H] as a function of $T_{\rm cond}$ by combining our measurements with those of \citet{Mack2014} using error-weighted means.
Applying the Monte Carlo approach described in Section~\ref{sec:sub:Tcond}, we find a negative correlation with $T_{\rm cond}$ across all measured elements, with a significance at the 2.7$\sigma$ level.
The integrated abundance difference of volatile elements, estimated from C, N, and O, is $\Delta$[X$_{\rm volatile}$/H] = 0.014~dex, whereas refractory elements with $T_{\rm cond} > 1300$~K show $\Delta$[X$_{\rm refractory}$/H] = $-0.054$~dex.
This indicates that refractory elements are more abundant in the atmosphere of HD~20781 than in HD~20782, while the volatile element abundances are similar in the two stars.

One possible interpretation of the observed trend is that it may reflect a planet-related signature, especially because this system hosts a combination of giant and low-mass planets broadly comparable to that of the Solar system.
The chemical signature of HD~20781, however, differs from that observed in the Sun, where refractory depletion has been suspected to be associated with the formation of nearby terrestrial planets, despite the presence of close-in low-mass planets around HD~20781.
If the observed abundance differences are primarily driven by planetary effects, the presence of close-in gas planets around HD~20781 may play a more significant role than low-mass planets in shaping the observed chemical signatures of the host stars.
On the other hand, it is also possible that refractory elements are depleted in HD~20782 as a consequence of pressure traps associated with the formation of a giant planet, which may inhibit the inward drift and accretion of refractory materials onto the host star \citep{Booth2020, Huhn2023}.
Whether such a mechanism can operate efficiently in the case of an extremely eccentric giant planet, such as HD~20782~b, remains unclear and requires further theoretical investigation.

In addition, an alternative explanation unrelated to planetary formation arises from the substantial differences in atmospheric parameters between the two stars, with HD~20782 being hotter by about 450~K and having a lower $\log{g}$ by approximately 0.2~dex than HD~20781 (see Table~\ref{tab:param}).
Such differences may lead to differential atomic diffusion, which can modify surface abundances and potentially contribute to the observed abundance differences \citep{Nordlander2012, Dotter2017, Liu2021}.


\subsection{WASP-160~A/WASP-160~B} \label{sec:sub:WASP-160}

In the WASP-160~A/B system, a gas giant planet has been discovered around the secondary star, WASP-160~B, while no planet has been reported around the primary component.
Using TESS photometry, \citet{Jofre2021} found a low probability for the presence of additional close-in planets in this system, although the existence of more distant or low-mass planets cannot be entirely excluded.
They also reported a statistically significant correlation between $\Delta$[X/H] and $T_{\rm cond}$, which qualitatively agrees with the trend observed in our analysis (Figure~\ref{fig:T_cond}).
As for the HD~20782/HD~20781 system, we combine our differential abundance measurements with those of \citet{Jofre2021}, excluding elements measured from a single spectral line.
The combined abundance differences are shown in the bottom panel of Figure~\ref{fig:T_cond_combine} as a function of $T_{\rm cond}$.
Although the resulting trend is dominated by the measurements of \citet{Jofre2021} owing to their smaller uncertainties, our data provide valuable constraints on the abundances of C, N, and O at the low-$T_{\rm cond}$ end, benefiting from the use of NIR spectroscopy.
We therefore regard the combined trend as a useful but non-independent comparison, rather than as a strict confirmation of our new measurements.
The combined data show a clear negative correlation with $T_{\rm cond}$ at high statistical significance.
This trend is qualitatively consistent with the trends found separately in our analysis and in that of \citet{Jofre2021}.
The integrated abundance differences are estimated to be $\Delta$[X$_{\rm volatile}$/H] = 0.128~dex and $\Delta$[X$_{\rm refractory}$/H] = $-0.039$~dex.
This indicates that the star hosting the close-in giant planet, WASP-160~B, is relatively enhanced in refractory elements and depleted in volatile elements compared to its wide companion without a detected planet, WASP-160~A.

For this system, the effect of atomic diffusion is expected to be less pronounced because of the small difference in atmospheric parameters between the two stars.
Several scenarios related to the presence of the gas giant planet are plausible, including the depletion of volatile material in WASP-160~B through sequestration of these materials into the gas giant during planet formation, or the enhancement of refractory elements via the accretion or ingestion of rocky material triggered by inward migration of the giant planet.
This interpretation is broadly consistent with the possible role of close-in gas planets discussed for the HD~20782/HD~20781 system.
We note that \citet{Jofre2021} also favored a planetary origin for the observed trend over non-planet-related explanations, although their quantitative analysis did not uniquely support any specific scenario related to planetary architecture.


\subsection{K2-54/K2-54~B} \label{sec:sub:K2-54}

This study presents the first high-resolution spectroscopic analysis of the K2-54/K2-54~B wide binary system.
As described in Section~\ref{sec:sub:Tcond}, this system shows no statistically significant chemical abundance trend with $T_{\rm cond}$, 
with $\Delta$[X$_{\rm volatile}$/H] = $-0.053$~dex and $\Delta$[X$_{\rm refractory}$/H] = $-0.010$~dex. 
This result indicates nearly identical chemical patterns in the two stellar components, despite the presence of a close-in gas planet orbiting K2-54.
Although a mildly positive slope is formally derived in Figure~\ref{fig:T_cond}, it is not significant compared to the scatter of the Monte Carlo realizations.
This result suggests that either K2-54~B may host as-yet undetected planets, leading to a similar chemical evolution in both components, or that the formation and subsequent evolution of the known planet around K2-54 did not produce a measurable chemical imprint on the host star.
Similar null results have been reported for other planet-hosting wide binaries, including HAT-P-1~A/B \citep{Liu2014} (see also Section~\ref{sec:sub:distribution} for additional examples).

The K2-54/K2-54~B system occupies a distinct region of parameter space within our sample.
It is the coolest pair and has the smallest projected separation (less than 1500~au) among the four wide binaries analyzed in this work, and the planet orbiting K2-54 (K2-54~b) is the lowest-mass planet among the gas planets considered here (see Table~\ref{tab:planet}).
These properties may contribute to the absence of a detectable $T_{\rm cond}$ trend, for example, by limiting the efficiency of planet-driven chemical imprinting on the stellar atmosphere or by diluting or preventing any potential signature during earlier evolutionary stages.
Taken together, the K2-54/K2-54~B system reinforces the conclusion that the presence of planets alone does not guarantee the emergence of a correlation between $\Delta$[X/H] and elemental $T_{\rm cond}$.
This motivates a broader investigation of how stellar, planetary, and binary properties influence chemical signatures, which we pursue in the following section using a larger sample of planet-hosting and non-planet-hosting wide binaries.


\subsection{WASP-127/TYC~4916-897-1}  \label{sec:sub:WASP-127}

The WASP-127/TYC~4916-897-1 system shows the strongest correlation between $\Delta$[X/H] and $T_{\rm cond}$ among our sample (Figure~\ref{fig:T_cond}).
As noted in Section~\ref{sec:sub:Tcond}, however, the sense of this trend differs from those observed in the HD~20782/HD~20781 and WASP-160~A/WASP-160~B systems, in that the planet-hosting star, WASP-127, is the primary component of the binary.
In this system, WASP-127 is significantly enhanced in volatile elements relative to its companion, with $\Delta$[X$_{\rm volatile}$/H] = 0.174~dex, while the two stars show nearly identical abundances of refractory elements ($\Delta$[X$_{\rm refractory}$/H] = 0.025~dex).
This abundance pattern is not straightforward to interpret in terms of planetary effects alone.
In particular, the presence of a close-in gas giant is generally expected either to reduce the volatile content of the host star through sequestration during planet formation or to affect refractory abundances by either promoting or inhibiting the accretion of rocky material, rather than selectively enhancing volatile elements.
One possible interpretation could invoke additional, as-yet undetected planets in the system, for example a close-in giant planet around TYC~4916-897-1, which might alter the relative abundance pattern.
However, such scenarios remain speculative and are not uniquely supported by the current observational constraints.

A more plausible explanation for the observed $\Delta$[X/H]-$T_{\rm cond}$ trend in this system is the different efficiencies of atomic diffusion between the two stellar components.
The difference in $T_{\rm eff}$ between the two components is 242~K, and the difference in $\log{g}$ is 0.30~dex, representing the largest atmospheric parameter mismatch among the planet-hosting wide binaries considered in this work.
When atmospheric parameters derived from grid-fitting are adopted, the difference in $\log{g}$ is reduced to 0.08~dex, but the difference in $T_{\rm eff}$ increases to 388~K.
Such differences are expected to lead to substantially different efficiencies of atomic diffusion between the two stars.
Therefore, atomic diffusion offers a plausible explanation for the strong $T_{\rm cond}$ trend observed in the WASP-127/TYC~4916-897-1 system, irrespective of the presence or absence of additional planets.
Although confirmation of this interpretation requires detailed stellar evolution modeling, as performed by \citet{Liu2024}, this system serves as a compelling example of how differential atomic diffusion can generate pronounced $\Delta$[X/H] trends with $T_{\rm cond}$ in wide binaries, even in the absence of clear planetary signatures.


\section{General trends of wide binaries in the literature} \label{sec:liter}

As discussed in the previous sections, the observed $T_{\rm cond}$ trends do not admit a unique interpretation, and their physical origin often remains ambiguous.
Although several observational studies, together with theoretical models, have explored possible star$-$planet interaction mechanisms in individual wide binary systems, firm conclusions have remained elusive \citep[e.g.,][]{Flores2024, Jofre2025}.
Moreover, some studies have argued that the chemical trends observed in the Sun and in planet-hosting stars are not directly associated with the presence of planetary systems \citep[e.g.,][]{Behmard2023, Ghezzi2026}.
These considerations motivate a statistical approach to assess whether the observed chemical abundance trends are more likely related to planetary architectures or to other stellar or binary properties.

To examine the relationship between binary $\Delta$[X/H]-$T_{\rm cond}$ trends and the presence of planets in a statistical sense, we compile high-resolution spectroscopic data for wide binary systems with and without reported planets from the literature.
Our sample consists of 15 planet-hosting wide binaries, listed in Table~\ref{tab:Tc_slope}, and 73 wide binaries without currently confirmed planets compiled from the literature \citep{Hawkins2020, Nagar2020, Liu2021, Nelson2021, Lim2024}, each with at least one measured abundance among C, N, and O.
In addition, we include three systems showing indirect evidence for planetary activity, such as chemical signatures of planet engulfment (HD~240430/HD~240429 and HIP~34407/HIP~34426) or the presence of a debris disk ($\zeta^{1}$~Ret/$\zeta^{2}$~Ret), despite the absence of confirmed planets \citep{Saffe2016, Oh2018, Nagar2020}.
Hereafter, for brevity, we refer to these categories as planet-hosting, non-planet-hosting, and candidate planet-hosting wide binaries.
It is important to note that this classification is based on the current observational status and may change as additional planets are discovered in systems currently classified as non-planet-hosting.
For systems with measurements from multiple sources, we compute error-weighted mean abundance differences for each element.
Using the same Monte Carlo approach described in Section~\ref{sec:sub:Tcond}, we estimate the slope of $\Delta$[X/H] as a function of $T_{\rm cond}$ (hereafter $T_{\rm cond}$ slope) and its 1$\sigma$ uncertainty.
For consistency across the compiled sample, we define $\Delta$[X/H] as the abundance of the planet-hosting (or more-planets) component minus that of the no-planet (or fewer-planets) component. 
This convention may therefore result in sign differences relative to earlier figures, where $\Delta$ was defined as primary minus secondary.
The resulting $T_{\rm cond}$ slopes and corresponding literature sources are summarized in Table~\ref{tab:Tc_slope}.
We note that our estimated slopes may differ from those reported in the original studies, as we exclude elements with large uncertainties and outliers relative to the overall trend, and adopt $T_{\rm cond}$ from \citet{Lodders2025}.

\begin{table*}
\centering
\caption{$T_{\rm cond}$ slopes for planet-hosting wide binary systems compiled from the literature and this work.}
\label{tab:Tc_slope} 
\begin{tabular}{ccccl}   
\hline\hline   
\multirow{2}{*}{System (A$-$B)} & \multirow{2}{*}{N$_{planet, A}$}    & \multirow{2}{*}{N$_{planet, B}$}
                                & $T_{\rm cond}$ slope $\pm~1\sigma$  & \multirow{2}{*}{Source} \\ 
                            & & & [10$^{-4}$ dex~K$^{-1}$]  & \\
\hline
\multicolumn{5}{c}{Planet-hosting systems} \\
\hline
WASP-160~B $-$ WASP-160~A       & 1 & 0 &    0.64 $\pm$ 0.06 & \citet{Jofre2021}, This~work \\
TOI-1173~A $-$ TOI-1173~B         & 1 & 0 &    0.61 $\pm$ 0.20 & \citet{YanaGalarza2024} \\
HD~20781 $-$ HD~20782           & 4 & 1 &    0.40 $\pm$ 0.15 & \citet{Mack2014}, This~work \\ 
HAT-P-4 $-$ TYC2569-744-1       & 1 & 0 &    0.38 $\pm$ 0.13 & \citet{Saffe2017} \\ 
HD~202772~A $-$ HD~202772~B     & 1 & 0 &    0.21 $\pm$ 0.08 & \citet{Jofre2025} \\ 
K2-54 $-$ K2-54~B               & 1 & 0 &    0.21 $\pm$ 0.27 & This~work \\ 
WASP-94~A $-$ WASP-94~B         & 1 & 1 &    0.20 $\pm$ 0.03 & \citet{Teske2016a} \\ 
16~$Cygni$~B $-$ 16~$Cygni$~A   & 1 & 0 &    0.07 $\pm$ 0.03 & \citet{Maia2019} \\ 
HAT-P-1~B $-$ HAT-P-1~A         & 1 & 0 &    0.03 $\pm$ 0.08 & \citet{Liu2014} \\
HD~106515~A $-$ HD~106515~B     & 1 & 0 & $-0.01$ $\pm$ 0.08 & \citet{Saffe2019}, \citet{Liu2021} \\
HD~133131~A $-$ HD~133131~B     & 2 & 1 & $-0.05$ $\pm$ 0.07 & \makecell[l]{\citet{Teske2016b}, \citet{Liu2021},\\ \citet{Nelson2021}} \\
HD~80606 $-$ HD~80607           & 1 & 0 & $-0.25$ $\pm$ 0.30 & \citet{Mack2016}, \citet{Hawkins2020} \\
HD~196067 $-$ HD~196068         & 2 & 0 & $-0.29$ $\pm$ 0.20 & \citet{Flores2024} \\
$XO$-2~S $-$ $XO$-2~N           & 2 & 1 & $-0.60$ $\pm$ 0.06 & \citet{Ramirez2015} \\
WASP-127 $-$ TYC~4916-897-1     & 1 & 0 & $-1.12$ $\pm$ 0.41 & This~work \\ 
\hline
\multicolumn{5}{c}{Candidate planet-hosting systems} \\
\hline
HD~240430 $-$ HD~240429         & 0 & 0 &    1.35 $\pm$ 0.14 & \citet{Oh2018}, \citet{Miquelarena2024} \\
HIP~34407 $-$ HIP~34426         & 0 & 0 &    0.77 $\pm$ 0.06 & \citet{Ramirez2019}, \citet{Nagar2020} \\
$\zeta^{1}~Ret$ $-$ $\zeta^{2}~Ret$ & 0 & 0 & 0.61 $\pm$ 0.20 & \citet{Saffe2016} \\
\hline
\end{tabular}
\tablefoot{The order A$-$B is defined such that component A hosts the same or a larger number of planets than component B.}
\end{table*}

\subsection{Comparison with planet-hosting and non-hosting wide binaries} \label{sec:sub:distribution}

\begin{figure}
\centering
   \includegraphics[width=0.48\textwidth]{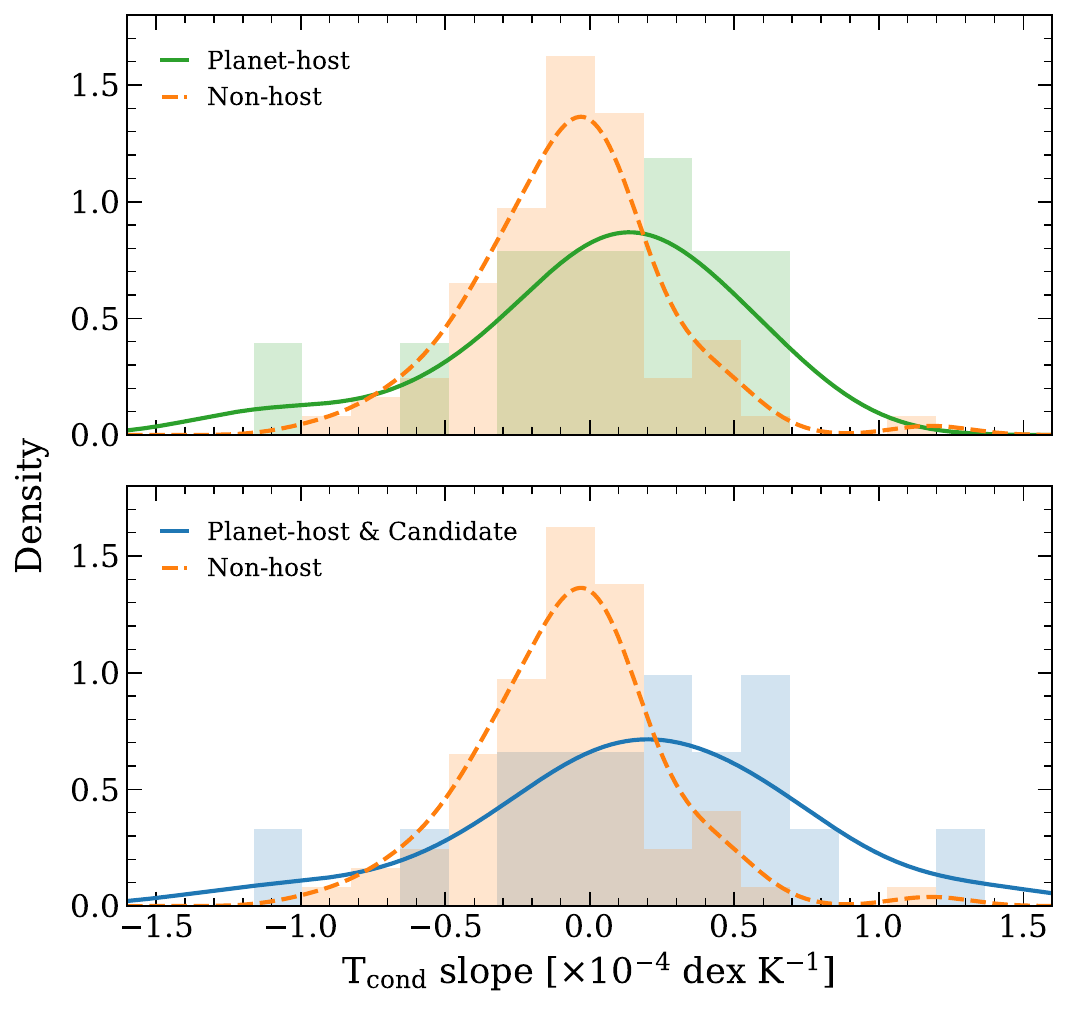}
     \caption{
     Comparison of the distributions of $T_{\rm cond}$ slopes for planet-hosting and non-hosting wide binaries.
     The histograms are normalized to unit area and use identical bin edges.
     Kernel density estimates are also plotted as solid lines for planet-hosting systems and dashed lines for non-hosting systems.
     In the upper panel, the green histogram represents the 15 confirmed planet-hosting wide binaries, while the orange histogram shows the non-hosting sample.
     In the lower panel, the planet-hosting sample is extended to include three candidate systems (blue), whereas the non-hosting sample remains unchanged.  
     Planet-hosting systems may exhibit a tentative excess in the tails of the distribution, although substantial overlap between the samples is present.
     }
     \label{fig:slope_hist}
\end{figure}

In Figure~\ref{fig:slope_hist}, we compare the distributions of $T_{\rm cond}$ slopes for planet-hosting and non-planet-hosting wide binary systems.
Although the two distributions largely overlap, planet-hosting systems show a tentative tendency to populate the extreme tails more frequently and to show a slight shift toward positive slopes.
A Kolmogorov-Smirnov (KS) test yields a $p$-value of 0.104, indicating that the null hypothesis that the two samples are drawn from the same parent distribution cannot be rejected.
In contrast, an Anderson-Darling (AD) test, which is more sensitive to differences in the tails of the distribution, yields a smaller $p$-value of 0.093.
Taken together, these results suggest that extreme $T_{\rm cond}$ slopes may be somewhat more common among planet-hosting wide binaries, although the current sample size does not allow a definitive conclusion.

When the candidate planet-hosting systems are included, the difference between the two distributions becomes more pronounced, as shown in the bottom panel of Figure~\ref{fig:slope_hist}.
In this case, the KS and AD tests yield $p$-values of 0.009 and 0.004, respectively.
This result suggests that the apparent difference between the distributions is sensitive to the inclusion of a small number of candidate systems.
Finally, we note that for non-planet-hosting systems the ordering of components (A$-$B or B$-$A) is arbitrary.
Repeating the KS and AD tests using the absolute values of the $T_{\rm cond}$ slopes yields larger $p$-values of 0.227 (KS) and 0.25 (AD) for the comparison with the planet-hosting sample, and 0.007 (KS) and 0.034 (AD) when the candidate systems are included.
This indicates that part of the observed difference may be influenced by sign conventions, while systems with confirmed or candidate planetary signatures may still tend to exhibit more extreme slopes.

On the other hand, \citet{Behmard2023} analyzed the chemical properties of a large sample of planet-hosting wide binaries based on Keck/HIRES spectroscopy and reported little evidence for planet engulfment in most of their systems.
For comparison, we estimated $T_{\rm cond}$ slopes using their published abundance measurements and examined the resulting distributions relative to those of non-hosting wide binaries.
We do not include these systems in our compiled sample, as the $T_{\rm cond}$ slopes derived from the \citet{Behmard2023} data show substantial offsets compared to those listed in Table~\ref{tab:Tc_slope}, with a mean difference of $\sim$0.35~dex per $10^{4}$~K for the 12 systems in common.
Consistent with the conclusions of \citet{Behmard2023}, the distribution of $T_{\rm cond}$ slopes inferred from their abundance measurements for planet-hosting wide binaries is statistically indistinguishable from that of non-hosting systems in our sample, yielding large $p$-values of 0.85 (KS test) and 0.25 (AD test).
This difference from our compiled sample indicates that the inferred $T_{\rm cond}$ slope distributions can depend substantially on the adopted abundance datasets and analysis choices.


\subsection{Dependence on binary separation} \label{sec:sub:separation}

\begin{figure}
\centering
   \includegraphics[width=0.48\textwidth]{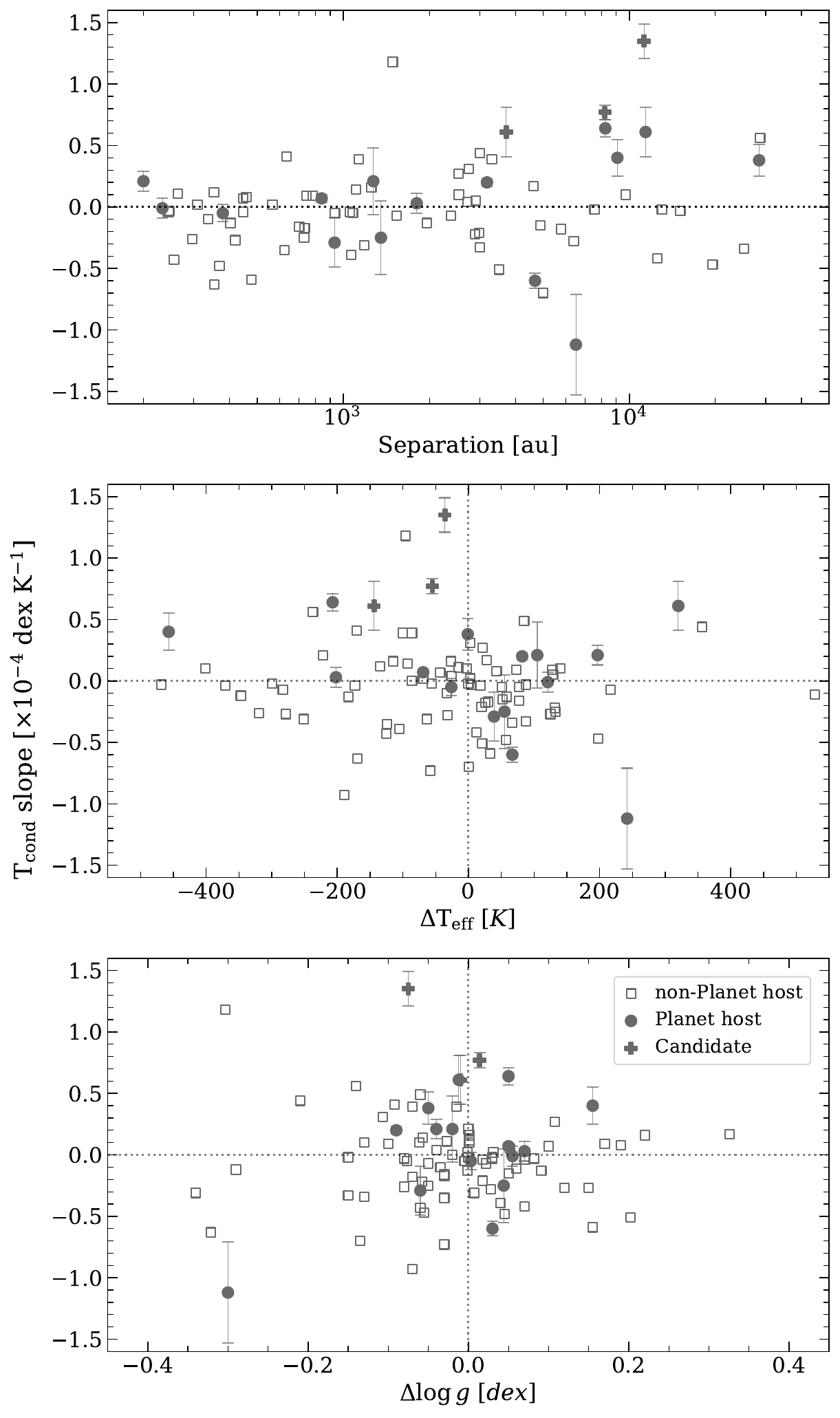}
     \caption{
     $T_{\rm cond}$ slopes as a function of projected binary separation (top), $\Delta T_{\rm eff}$ (middle), and $\Delta \log{g}$ (bottom) between the two components of each system.
     Planet-hosting wide binaries and candidate systems are shown as filled circles and plus symbols, respectively, with vertical error bars indicating the $\pm1\sigma$ uncertainty of the slope, while non-planet-hosting systems are shown as open squares.
     The horizontal dotted line marks zero slope, and the vertical lines in the middle and bottom panels indicate zero difference in $T_{\rm eff}$ and $\log{g}$.
     One remarkable feature is that extreme $T_{\rm cond}$ slopes are preferentially found at large separations in the planet-hosting sample.
     }
     \label{fig:slope_trend}
\end{figure}

To assess whether the observed positive or negative $T_{\rm cond}$ slopes could arise from effects other than planetary signatures, we examine the dependence of $T_{\rm cond}$ slopes on projected separation, $\Delta T_{\rm eff}$, and $\Delta \log g$, as shown in Figure~\ref{fig:slope_trend}.
With respect to binary separation, it has been suggested that wide binaries formed from larger and less chemically homogeneous gas clouds may exhibit enhanced abundance differences with increasing separation, in an element-dependent manner \citep{Ramirez2019, Lim2024}.
However, separation-driven abundance differences alone are not expected to produce a systematic correlation between $\Delta$[X/H] and $T_{\rm cond}$.
Consistent with this expectation, we find no correlation between $T_{\rm cond}$ slope and separation for non-planet-hosting wide binaries (upper panel of Figure~\ref{fig:slope_trend}), indicating that separation alone cannot account for systematic variations in $T_{\rm cond}$ trends.

For planet-hosting wide binaries, a different behavior emerges in the dependence of $T_{\rm cond}$ slope on separation.
Systems with projected separations smaller than 2000~au tend to show $T_{\rm cond}$ slopes clustered around zero, whereas those with larger separations ($>$ 2000~au) display a substantially broader distribution, spanning both positive and negative values.
The standard deviation of the $T_{\rm cond}$ slope for planet-hosting systems increases from 0.18 (8 systems) at separations $<$ 2000~au to 0.62 (7 systems) at separations $>$ 2000~au, while the corresponding values for non-hosting binaries remain nearly constant at 0.32 (35 systems) and 0.33 (38 systems), respectively.
When the planet-hosting sample is further divided into three separation bins ($<$ 600~au, 600–2000~au, and $>$ 2000~au), the scatter in $T_{\rm cond}$ slope increases with separation, from 0.114 to 0.193 and 0.621, respectively.
We emphasize that these separation boundaries are empirical and adopted for illustrative purposes, and that the true three-dimensional separations may be substantially larger than the projected values.

Our results do not support a purely separation-driven origin of the observed $T_{\rm cond}$ trends, as non-hosting systems exhibit nearly identical slope distributions at both small and large separations.
Instead, a more plausible interpretation is that the presence of a stellar companion at separations up to $\sim$2000~au may modulate the ability of planetary systems to imprint chemical signatures on their host stars.
This interpretation is consistent with previous studies showing that planet occurrence rates and planetary architectures differ between single stars and stars in binary or multiple systems \citep[e.g.,][]{Wang2014, Winter2020, Hirsch2021}.
One possible explanation is that wide binaries with relatively small separations may have experienced stronger early dynamical or chemical interactions, which could dilute or erase planet-induced chemical signatures in stellar atmospheres.
In contrast, systems with larger separations may have evolved more independently from an early stage, allowing chemical abundance patterns associated with planet formation and evolution to be preserved in the host stars.
Furthermore, different formation pathways for wide binaries, such as formation from adjacent pre-stellar cores \citep{Tokovinin2017} or the dynamical unfolding of higher-order multiple systems \citep{Reipurth2012}, may naturally lead to diverse planetary environments and chemical outcomes.
If different formation channels preferentially dominate at smaller versus larger binary separations, this could in turn influence the likelihood that planetary chemical signatures are preserved or erased in the host stellar atmospheres. 
In any case, a detailed theoretical treatment is required to identify the physical mechanisms underlying the observed discrepancy between closer and wider binaries.
Nevertheless, it is notable that the strongest $\Delta$[X/H]–$T_{\rm cond}$ correlations within the current compiled sample are preferentially found among planet-hosting wide binaries with separations larger than 2000~au.

\subsection{The role of differential atomic diffusion} \label{sec:sub:diffusion}

The middle and bottom panels of Figure~\ref{fig:slope_trend} illustrate the dependence of $T_{\rm cond}$ slope on differences in atmospheric parameters between the two components of each wide binary system.
Several studies have suggested that atomic diffusion operating in binaries with different $T_{\rm eff}$ or $\log{g}$ can produce abundance trends with $T_{\rm cond}$, even in the absence of planetary effects \citep[e.g.,][]{Liu2021, Liu2024}.
The abundance differences induced by atomic diffusion are typically of order 0.01 to 0.05~dex for solar-type stars \citep[e.g.,][]{Dotter2017}, although the exact magnitude depends on stellar mass and evolutionary stage \citep[e.g.,][]{Korn2007}. 
Such effects are expected to be more pronounced in systems with large differences in $T_{\rm eff}$ and/or $\log{g}$.
However, no clear correlation is observed between $T_{\rm cond}$ slope and either $\Delta T_{\rm eff}$ or $\Delta \log{g}$ across the full sample shown in Figure~\ref{fig:slope_trend}, suggesting that atomic diffusion alone does not provide a general explanation for the observed $\Delta$[X/H]-$T_{\rm cond}$ trends.
Two systems located at the extremes of the parameter space, HD~20781$-$HD~20782 ($\Delta T_{\rm eff} \simeq -450$~K, $\Delta \log{g} \simeq 0.2$~dex) and WASP-127$-$TYC~4916-897-1 ($\Delta T_{\rm eff} \simeq 240$~K, $\Delta \log{g} \simeq -0.3$~dex), exhibit particularly strong $T_{\rm cond}$ slopes, and atomic diffusion may contribute to the observed trends in these cases.
Nevertheless, several systems with small differences in both $T_{\rm eff}$ and $\log{g}$ still display pronounced positive or negative $T_{\rm cond}$ slopes, indicating that additional processes are likely at play.

To further assess the role of atomic diffusion, we restrict the sample to systems with relatively small differences in atmospheric parameters.
Adopting a practical threshold of $|\Delta T_{\rm eff}| < 250$~K and $|\Delta \log{g}| < 0.15$~dex, which is expected to substantially reduce, though not eliminate, differential atomic diffusion effects, we retain 12 planet-hosting and three candidate planet-hosting wide binaries.
Within this restricted sample, five systems exhibit positive ($T_{\rm cond}$ slope $\ge 0.2$) trends and three show negative ($T_{\rm cond}$ slope $\le -0.2$) trends, while the remaining four systems display nearly flat relations ($-0.1 < T_{\rm cond}$ slope $< +0.1$).
A comparison with the 51 non-planet-hosting systems satisfying the same criteria yields $p$-values of 0.449 (KS test) and 0.25 (AD test), indicating no statistically significant difference.
When the three candidate planet-hosting systems are included, the $p$-values decrease to 0.056 (KS) and 0.016 (AD), suggesting a possible excess of extreme $T_{\rm cond}$ slopes among systems with confirmed or suspected planetary signatures, although the result remains sensitive to small-number statistics.

We further confirm that the increased scatter in $T_{\rm cond}$ slope observed at large binary separations (top panel of Figure~\ref{fig:slope_trend}) does not appear to be driven by systematic trends in either $\Delta T_{\rm eff}$ or $\Delta \log{g}$ with separation.
No clear correlation with separation is observed for either planet-hosting or non-hosting samples, although a small number of extreme systems at separations larger than 2000~au are excluded by the above cuts.


\section{Summary and Discussion} \label{sec:discussion}

In this study, we conducted high-resolution NIR spectroscopy of four planet-hosting wide binary systems using IGRINS at the Gemini-South Observatory.
Atmospheric parameters derived from photometric and grid-fitting approaches show generally good agreement in $T_{\rm eff}$, while modest systematic differences are found in $\log{g}$.
Using the photometric parameter set, we measure small chemical abundance differences between the binary components, with mean absolute differences of $\overline{|\Delta\mathrm{[X/H]}|}$ $\simeq$ 0.02$-$0.08~dex.
In addition, two systems, WASP-160~A/WASP-160~B and WASP-127/TYC~4916-897-1, exhibit statistically significant correlations between $\Delta$[X/H] and $T_{\rm cond}$.
Notably, these two systems show opposite behaviors as volatile elements are relatively depleted in the planet-hosting star WASP-160~B, whereas they are enhanced in the planet-hosting star WASP-127.
The HD~20782/HD~20781 system shows only a marginal correlation, while the K2-54/K2-54~B system is consistent with a flat relation.
These results indicate that chemical patterns observed in planet-hosting wide binaries are diverse from system to system.

Our system-by-system analysis suggests the following possible interpretations.
For HD~20782/HD~20781, the observed abundance pattern may be related to the different planetary architectures of the two stars, although alternative explanations such as differential atomic diffusion remain plausible.
In the WASP-160~A/WASP-160~B system, an enhancement of refractory elements, or depletion of volatile elements, is consistent with the gas giant orbiting WASP-160~B.
In contrast, the absence of a detectable chemical signature in K2-54/K2-54~B may be related to the relatively small binary separation or the low mass of the known planet.
Finally, for WASP-127/TYC~4916-897-1, the strong $\Delta$[X/H]--$T_{\rm cond}$ trend is more plausibly explained by differential atomic diffusion between the two components.

Extending our analysis to a larger sample from the literature, we find that the distribution of $T_{\rm cond}$ slopes differs modestly between planet-hosting and non-planet-hosting wide binaries.
This difference becomes more pronounced when systems with indirect evidence for planetary activity are included.
In addition, we find that strong $T_{\rm cond}$ slopes are preferentially observed among planet-hosting wide binaries with projected separations larger than $\sim$2000~au.
This trend may reflect differences in the formation and early dynamical evolution of wide binary systems, which in turn could modulate the ability of planetary systems to imprint chemical signatures on their host stars.

A key distinction of our study relative to previous investigations of planet-hosting wide binaries is the use of high-resolution NIR spectroscopy.
Compared to optical spectroscopy, the NIR wavelength range allows more reliable abundance determinations for volatile elements based on numerous molecular and atomic features \citep{Lim2024}.
In addition, the observational feasibility of NIR spectroscopy for cool, planet-hosting stars represents a further advantage of our approach.
On the other hand, the more limited set of measurable elements in the NIR regime constitutes a potential limitation.
In this study, although chemical abundances for 18 elements were measured, only 13 elements provide sufficiently precise measurements for detailed analyses of planet-hosting stars.
As a result, $T_{\rm cond}$ slopes derived solely from our NIR measurements exhibit relatively large scatter, as shown in Figure~\ref{fig:T_cond} and Table~\ref{tab:Tc_slope}.
Taken together, these strengths and limitations highlight the complementarity of optical and NIR spectroscopy, and suggest that a combined approach would be particularly powerful for future studies of chemical signatures in planet-hosting stars.

In Section~\ref{sec:liter}, we performed a collective analysis of wide binary systems with and without confirmed planets.
Such an approach is essential given the ongoing debate over whether differential abundance trends with $T_{\rm cond}$ can be uniquely attributed to planetary signatures.
While the present analysis is based on a limited sample and should therefore be regarded as suggestive rather than conclusive, this primarily highlights the need for future extensions rather than undermining the tentative trends identified here.
The statistical significance of the inferred trends increases when three candidate systems are included, indicating that additional well-characterized samples will be important for testing whether these tentative trends remain robust.
We also note that a fraction of systems classified as non-planet-hosting may in fact harbor undetected planets around one or both components.
Such hidden planetary systems would effectively dilute the contrast between planet-hosting and non-hosting samples, potentially reducing the statistical significance of any intrinsic differences.

In this context, the classification adopted for individual wide binary systems should be understood as distinguishing stars hosting at least one confirmed close-in planet from stars with a low probability of hosting nearby planets, rather than as a strict dichotomy between planet-hosting and non-hosting stars.
Undetected distant or low-mass planets may exist around either component.
We expect, however, that the chemical impact of such planets on the host stellar atmosphere would be smaller than that of the currently known planetary systems, since close-in and massive planets are both more likely to induce detectable chemical signatures and to be identified observationally.
It is also possible that some close-in planets remain undetected due to unfavorable orbital inclinations.
Future $Gaia$ data releases, particularly those including epoch astrometry, may help to identify such systems \citep[see][]{Wallace2025} and improve the census of planets in wide binary systems.

\begin{figure}
\centering
   \includegraphics[width=0.35\textwidth]{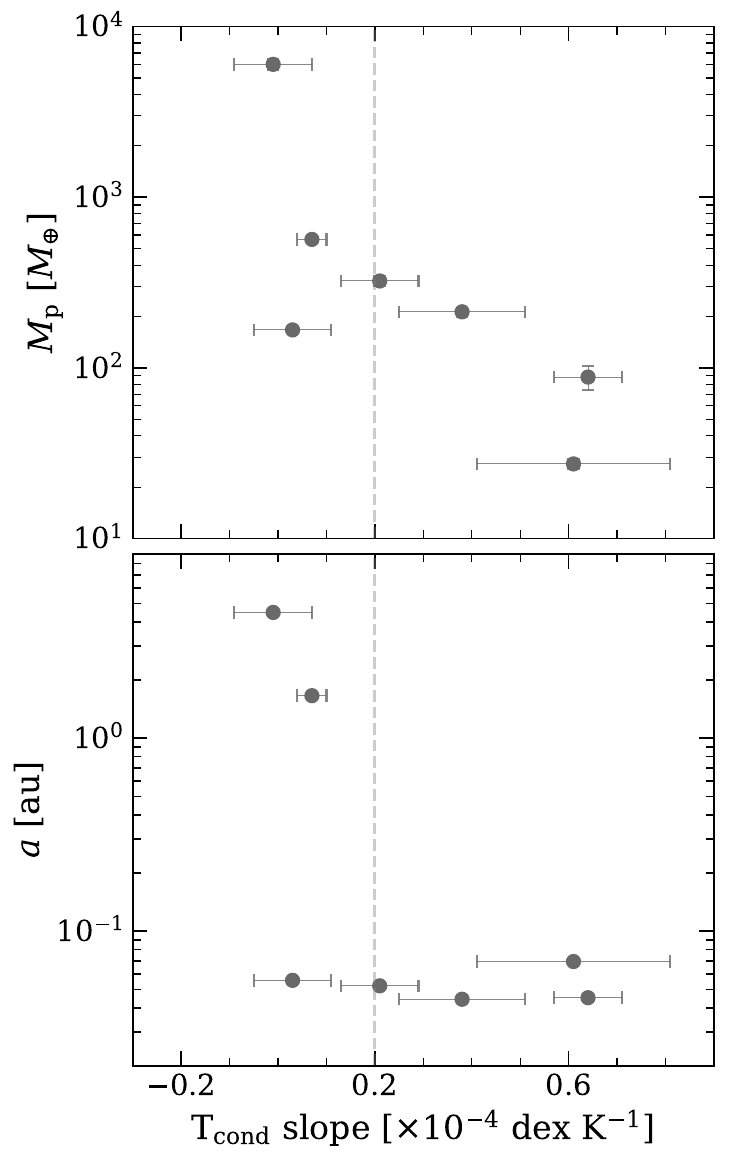}
     \caption{
     Planetary mass (top) and semi-major axis (bottom) as functions of the $T_{\rm cond}$ slope for a subset of seven planet-hosting wide binaries.
     The $\pm1\sigma$ uncertainties of the $T_{\rm cond}$ slopes and the measurement errors in planetary mass and semi-major axis are indicated, respectively, although the latter are typically smaller than the symbol size.
     The vertical dashed line at $T_{\rm cond}$ slope = 0.2 separates systems with positive and near-zero $\Delta$[X/H]–$T_{\rm cond}$ trends.
     While systems with positive slopes tend to host planets of moderate mass, this figure is intended to illustrate possible patterns within a limited sample and does not imply a statistically significant correlation.
     }
     \label{fig:slope_planet}
\end{figure}

Furthermore, we briefly explore whether the $T_{\rm cond}$ slope may be related to basic planetary properties, while emphasizing that this analysis is highly limited by small-number statistics.
To minimize potential confounding effects, we restrict this analysis to seven planet-hosting wide binaries, excluding multi-planetary systems, systems likely affected by strong differential atomic diffusion, and systems with large uncertainties in the derived $T_{\rm cond}$ slopes.
Within this reduced sample, four systems exhibit positive $T_{\rm cond}$ slopes (WASP-160~B/WASP-160~A, TOI-1173~A/TOI-1173~B, HAT-P-4/TYC~2569-744-1, and HD~202772~A/HD~202772~B), while the remaining three systems show slopes consistent with zero (16~Cyg~B/16~Cyg~A, HAT-P-1~B/HAT-P-1~A, and HD~106515~A/HD~106515~B).
We find a tentative indication that the magnitude of the $T_{\rm cond}$ slope may weakly anti-correlate with planetary mass (see Figure~\ref{fig:slope_planet}).
Systems with positive $T_{\rm cond}$ slopes tend to host planets of moderate mass (approximately 30–300~$M_{\oplus}$) on close-in orbits ($a \sim 0.05$~au), whereas systems with near-zero slopes host more massive and/or more distant planets, such as 16~Cyg~B~b and HD~106515~A~b.
We caution that notable exceptions exist, most prominently HAT-P-1~B/HAT-P-1~A, which hosts a planet with comparable mass and semi-major axis to systems showing positive slopes, yet exhibits no significant $T_{\rm cond}$ trend.
Given the small sample size and the absence of a clear theoretical framework predicting such behavior, we refrain from drawing firm conclusions.
Nevertheless, this tentative pattern may hint at a threshold in planetary mass or orbital configuration required to imprint detectable chemical signatures on host stars, a possibility that warrants further investigation with larger samples and dedicated theoretical modeling.

Further progress in understanding $T_{\rm cond}$ trends in wide binaries, both with and without planets, will require larger samples with well-characterized planetary architectures.
Such samples should include not only confirmed planet-hosting systems but also wide binaries classified as non-planet-hosting, enabling more robust comparisons and a clearer separation of planetary signatures from stellar and binary effects.
In addition, incorporating stellar properties such as age, mass, and evolutionary stage, which were not explicitly considered in this study, for both planet-hosting and non-hosting systems, will help establish a more comprehensive understanding of how planetary and stellar effects together shape the observed chemical patterns in wide binary systems.


\begin{acknowledgements}
We are grateful to the referee for a number of helpful suggestions.
D.L. acknowledges support from Basic Science Research Program through the National Research Foundation (NRF) of Korea funded by the Ministry of Education (RS-2025-25419201 and RS-2022-NR070872).
Y.S.L. acknowledges support from the NRF of Korea grant funded by the Ministry of Science and ICT (RS-2024-00333766).
D.L. thanks Sree Oh for the ongoing support.
This work was supported by K-GMT Science Program (PID: GS-2023B-Q-211 and GS-2023B-Q-315) of Korea Astronomy and Space Science Institute (KASI). 
This work used the Immersion Grating Infrared Spectrometer (IGRINS) that was developed under a collaboration between the University of Texas at Austin and the KASI with the financial support of the Mt. Cuba Astronomical Foundation, of the US National Science Foundation under grants AST-1229522 and AST-1702267, of the McDonald Observatory of the University of Texas at Austin, of the Korean GMT Project of KASI, and Gemini Observatory. 
\end{acknowledgements}

%
   \bibliographystyle{aa} 
   \bibliography{export-bibtex} 

\onecolumn
\begin{appendix}

\section{Full abundance table} \label{app:abund}

We provide the full elemental abundance tables derived using both photometric and grid-fitting atmospheric parameters for all four wide binary systems analyzed in this work.

\begin{table}[!ht]
\centering
\small 
\renewcommand{\arraystretch}{0.99}
\caption{
Elemental abundance ratios [X/H] and uncertainties for HD~20782 and HD~20781 derived using both photometric and grid-fitting atmospheric parameters.
}
\label{tab:abund_HD20782}
\begin{tabular}{@{\extracolsep{4pt}}lrrrr}
\hline\hline
\multirow{2}{*}{[X/H]} & \multicolumn{2}{c}{Photometric parameters} & \multicolumn{2}{c}{Grid-fitting parameters} \\
\cline{2-3} \cline{4-5}
   & HD~20782              & HD~20781              & HD~20782              & HD~20781 \\
\hline
Fe & $-0.06 \pm 0.02$ (26) & $ 0.04 \pm 0.02$ (24) & $-0.10 \pm 0.01$ (25) & $ 0.04 \pm 0.02$ (24) \\
C  & $-0.22 \pm 0.02$ (15) & $-0.04 \pm 0.01$ (15) & $-0.38 \pm 0.02$ (15) & $-0.02 \pm 0.01$ (15) \\
N  & $-0.06 \pm 0.04$ (10) & $-0.12 \pm 0.03$ (15) & $-0.02 \pm 0.04$ (10) & $-0.16 \pm 0.03$ (15) \\
O  & $ 0.17 \pm 0.07$ (4)  & $ 0.06 \pm 0.07$ (10) & $ 0.15 \pm 0.07$ (2)  & $ 0.04 \pm 0.07$ (10) \\
Na & $-0.09 \pm 0.02$ (3)  & $ 0.03 \pm 0.01$ (3)  & $-0.18 \pm 0.02$ (3)  & $-0.03 \pm 0.01$ (3)  \\
Mg & $-0.17 \pm 0.04$ (10) & $-0.11 \pm 0.04$ (10) & $-0.26 \pm 0.06$ (10) & $-0.15 \pm 0.05$ (10) \\
Al & $ 0.09 \pm 0.08$ (4)  & $ 0.19 \pm 0.08$ (4)  & $ 0.04 \pm 0.09$ (4)  & $ 0.17 \pm 0.08$ (4)  \\
Si & $-0.05 \pm 0.02$ (12) & $-0.01 \pm 0.02$ (12) & $-0.09 \pm 0.03$ (12) & $-0.01 \pm 0.02$ (12) \\
P  & $-0.09 \pm 0.15$ (2)  & \ldots                & $-0.02 \pm 0.15$ (2)  & \ldots                \\
S  & $ 0.05 \pm 0.02$ (9)  & $ 0.11 \pm 0.03$ (9)  & $ 0.14 \pm 0.02$ (9)  & $ 0.17 \pm 0.03$ (9)  \\
K  & $-0.04 \pm 0.00$ (2)  & $-0.05 \pm 0.01$ (2)  & $-0.07 \pm 0.00$ (2)  & $-0.07 \pm 0.02$ (2)  \\
Ca & $-0.02 \pm 0.01$ (9)  & $ 0.00 \pm 0.01$ (9)  & $-0.07 \pm 0.02$ (9)  & $-0.02 \pm 0.02$ (9)  \\
Sc & $-0.04 \pm 0.02$ (2)  & $ 0.03 \pm 0.00$ (1)  & $-0.14 \pm 0.04$ (2)  & $-0.01 \pm 0.01$ (2)  \\
Ti & $-0.18 \pm 0.03$ (8)  & $-0.04 \pm 0.03$ (8)  & $-0.26 \pm 0.03$ (8)  & $-0.06 \pm 0.03$ (8)  \\
Cr & $-0.04 \pm 0.02$ (2)  & $-0.06 \pm 0.01$ (2)  & $-0.08 \pm 0.02$ (2)  & $-0.06 \pm 0.01$ (2)  \\
Co & $-0.52 \pm \ldots$ (1)& $-0.33 \pm \ldots$ (1)& $-0.57 \pm \ldots$ (1)& $-0.28 \pm \ldots$ (1)\\
Ni & $-0.06 \pm 0.02$ (5)  & $-0.00 \pm 0.02$ (5)  & $-0.08 \pm 0.02$ (5)  & $ 0.01 \pm 0.02$ (5)  \\
Yb & $-0.22 \pm \ldots$ (1)& $-0.13 \pm \ldots$ (1)& $-0.16 \pm \ldots$ (1)& $-0.08 \pm \ldots$ (1)\\
\hline
\end{tabular}
\tablefoot{Numbers in parentheses indicate the number of lines used for each element.
Uncertainties correspond to the standard error of the mean.
For elements measured from a single line, the standard error cannot be estimated and is therefore not reported.}
\end{table}

\begin{table}[!ht]
\centering
\small 
\renewcommand{\arraystretch}{0.99}
\caption{
Same as Table~\ref{tab:abund_HD20782} but for WASP-160~A and WASP-160~B.
}
\label{tab:abund_WASP160}
\begin{tabular}{@{\extracolsep{4pt}}lrrrr}
\hline\hline
\multirow{2}{*}{[X/H]} & \multicolumn{2}{c}{Photometric parameters} & \multicolumn{2}{c}{Grid-fitting parameters} \\
\cline{2-3} \cline{4-5}
   & WASP-160~A            & WASP-160~B            & WASP-160~A            & WASP-160~B \\
\hline
Fe & $ 0.20 \pm 0.02$ (26) & $ 0.22 \pm 0.03$ (27) & $ 0.15 \pm 0.02$ (26) & $ 0.15 \pm 0.03$ (27) \\
C  & $-0.03 \pm 0.01$ (15) & $ 0.08 \pm 0.01$ (16) & $-0.10 \pm 0.01$ (15) & $-0.06 \pm 0.01$ (16) \\
N  & $ 0.61 \pm 0.02$ (16) & $ 0.38 \pm 0.02$ (15) & $ 0.50 \pm 0.02$ (16) & $ 0.26 \pm 0.02$ (15) \\
O  & $ 0.47 \pm 0.02$ (4)  & $ 0.23 \pm 0.04$ (8)  & $ 0.29 \pm 0.03$ (4)  & $ 0.01 \pm 0.04$ (8)  \\
Na & $ 0.30 \pm 0.01$ (3)  & $ 0.42 \pm 0.00$ (3)  & $ 0.19 \pm 0.01$ (3)  & $ 0.28 \pm 0.00$ (3)  \\
Mg & $ 0.02 \pm 0.04$ (10) & $ 0.02 \pm 0.04$ (10) & $-0.05 \pm 0.05$ (10) & $-0.07 \pm 0.05$ (10) \\
Al & $ 0.31 \pm 0.07$ (6)  & $ 0.35 \pm 0.07$ (6)  & $ 0.24 \pm 0.07$ (6)  & $ 0.26 \pm 0.07$ (6)  \\
Si & $ 0.18 \pm 0.03$ (12) & $ 0.19 \pm 0.03$ (12) & $ 0.15 \pm 0.03$ (12) & $ 0.17 \pm 0.03$ (12) \\
P  & $ 0.28 \pm \ldots$ (1)& $ 0.65 \pm \ldots$ (1)& $ 0.32 \pm \ldots$ (1)& $ 0.68 \pm \ldots$ (1)\\
S  & $ 0.21 \pm 0.04$ (5)  & $ 0.24 \pm 0.05$ (5)  & $ 0.28 \pm 0.03$ (5)  & $ 0.29 \pm 0.05$ (5)  \\
K  & $ 0.15 \pm 0.03$ (2)  & $ 0.21 \pm 0.00$ (2)  & $ 0.09 \pm 0.04$ (2)  & $ 0.13 \pm 0.01$ (2)  \\
Ca & $ 0.19 \pm 0.03$ (10) & $ 0.23 \pm 0.03$ (10) & $ 0.11 \pm 0.02$ (10) & $ 0.12 \pm 0.03$ (10) \\
Sc & $ 0.23 \pm \ldots$ (1)& $-0.02 \pm \ldots$ (1)& $ 0.11 \pm \ldots$ (1)& $-0.21 \pm \ldots$ (1)\\
Ti & $ 0.20 \pm 0.04$ (8)  & $ 0.21 \pm 0.05$ (8)  & $ 0.10 \pm 0.04$ (8)  & $ 0.05 \pm 0.04$ (8)  \\
Cr & $ 0.06 \pm 0.05$ (3)  & $ 0.09 \pm 0.05$ (3)  & $ 0.01 \pm 0.05$ (3)  & $-0.00 \pm 0.05$ (3)  \\
Co & $-0.20 \pm \ldots$ (1)& $ 0.01 \pm \ldots$ (1)& $-0.24 \pm \ldots$ (1)& $-0.06 \pm \ldots$ (1)\\
Ni & $ 0.25 \pm 0.01$ (5)  & $ 0.27 \pm 0.02$ (5)  & $ 0.23 \pm 0.02$ (5)  & $ 0.24 \pm 0.02$ (5)  \\
Yb & $ 0.11 \pm \ldots$ (1)& $ 0.59 \pm \ldots$ (1)& $ 0.13 \pm \ldots$ (1)& $ 0.57 \pm \ldots$ (1)\\
\hline
\end{tabular}
\end{table}

\begin{table}[!ht]
\centering
\small 
\renewcommand{\arraystretch}{0.99}
\caption{
Same as Table~\ref{tab:abund_HD20782} but for K2-54 and K2-54~B.
}
\label{tab:abund_K2_54}
\begin{tabular}{@{\extracolsep{4pt}}lrrrr}
\hline\hline
\multirow{2}{*}{[X/H]} & \multicolumn{2}{c}{Photometric parameters} & \multicolumn{2}{c}{Grid-fitting parameters} \\
\cline{2-3} \cline{4-5}
   & K2-54                 & K2-54~B               & K2-54                 & K2-54~B \\
\hline
Fe & $ 0.06 \pm 0.02$ (21) & $ 0.06 \pm 0.03$ (22) & $ 0.12 \pm 0.01$ (21) & $ 0.13 \pm 0.02$ (23) \\
C  & $-0.22 \pm 0.02$ (16) & $-0.22 \pm 0.02$ (16) & $ 0.00 \pm 0.02$ (16) & $ 0.02 \pm 0.02$ (16) \\
N  & $ 0.02 \pm 0.04$ (15) & $ 0.12 \pm 0.06$ (15) & $ 0.12 \pm 0.04$ (15) & $ 0.18 \pm 0.05$ (15) \\
O  & $-0.27 \pm 0.01$ (12) & $-0.26 \pm 0.01$ (12) & $-0.01 \pm 0.01$ (12) & $ 0.01 \pm 0.01$ (12) \\
Na & $ 0.13 \pm 0.06$ (3)  & $ 0.12 \pm 0.03$ (3)  & $ 0.24 \pm 0.07$ (3)  & $ 0.19 \pm 0.05$ (3)  \\
Mg & $-0.16 \pm 0.03$ (7)  & $-0.08 \pm 0.03$ (8)  & $-0.22 \pm 0.05$ (7)  & $-0.18 \pm 0.06$ (7)  \\
Al & $ 0.13 \pm 0.07$ (5)  & $ 0.12 \pm 0.07$ (5)  & $ 0.21 \pm 0.08$ (5)  & $ 0.18 \pm 0.08$ (5)  \\
Si & $-0.02 \pm 0.02$ (10) & $-0.02 \pm 0.02$ (10) & $ 0.01 \pm 0.02$ (10) & $ 0.00 \pm 0.02$ (10) \\
P  & \ldots                & \ldots                & \ldots                & \ldots \\
S  & $ 0.87 \pm 0.05$ (5)  & \ldots                & $ 0.75 \pm 0.07$ (5)  & \ldots \\
K  & $-0.03 \pm 0.01$ (2)  & $-0.03 \pm 0.02$ (2)  & $ 0.04 \pm 0.00$ (2)  & $ 0.04 \pm 0.01$ (2)  \\
Ca & $-0.02 \pm 0.02$ (9)  & $-0.03 \pm 0.02$ (9)  & $ 0.08 \pm 0.03$ (9)  & $ 0.06 \pm 0.03$ (9)  \\
Sc & $-0.02 \pm 0.06$ (2)  & $-0.10 \pm 0.04$ (2)  & $ 0.20 \pm 0.07$ (2)  & $ 0.12 \pm 0.04$ (2)  \\
Ti & $ 0.14 \pm 0.04$ (9)  & $ 0.15 \pm 0.03$ (9)  & $ 0.28 \pm 0.04$ (9)  & $ 0.29 \pm 0.03$ (9)  \\
Cr & $-0.06 \pm 0.04$ (3)  & $-0.08 \pm 0.04$ (3)  & $ 0.04 \pm 0.03$ (3)  & $ 0.03 \pm 0.02$ (3)  \\
Co & $-0.17 \pm \ldots$ (1)& $-0.16 \pm \ldots$ (1)& $-0.04 \pm \ldots$ (1)& $-0.02 \pm \ldots$ (1)\\
Ni & $ 0.04 \pm 0.01$ (3)  & $ 0.06 \pm 0.02$ (3)  & $ 0.12 \pm 0.01$ (3)  & $ 0.13 \pm 0.03$ (3)  \\
Yb & \ldots                & $ 0.54 \pm \ldots$ (1)& \ldots                & $ 0.67 \pm \ldots$ (1)\\
\hline
\end{tabular}
\end{table}

\begin{table}[!ht]
\centering
\small 
\renewcommand{\arraystretch}{0.99}
\caption{
Same as Table~\ref{tab:abund_HD20782} but for WASP-127 and TYC~4916-897-1.
}
\label{tab:abund_WASP127}
\begin{tabular}{@{\extracolsep{4pt}}lrrrr}
\hline\hline
\multirow{2}{*}{[X/H]} & \multicolumn{2}{c}{Photometric parameters} & \multicolumn{2}{c}{Grid-fitting parameters} \\
\cline{2-3} \cline{4-5}
   & WASP-127              & TYC~4916-897-1        & WASP-127              & TYC~4916-897-1 \\
\hline
Fe & $-0.14 \pm 0.04$ (20) & $-0.19 \pm 0.02$ (25) & $-0.10 \pm 0.04$ (20) & $-0.22 \pm 0.01$ (24) \\
C  & $-0.16 \pm 0.04$ (10) & $-0.28 \pm 0.03$ (16) & $-0.01 \pm 0.04$ (10) & $-0.34 \pm 0.03$ (16) \\
N  & $ 0.02 \pm 0.09$ (6)  & $-0.22 \pm 0.04$ (12) & $-0.01 \pm 0.08$ (6)  & $-0.29 \pm 0.03$ (12) \\
O  & $ 0.21 \pm 0.09$ (7)  & $ 0.05 \pm 0.06$ (7)  & $ 0.23 \pm 0.10$ (7)  & $-0.06 \pm 0.05$ (7)  \\
Na & $-0.07 \pm 0.06$ (4)  & $-0.24 \pm 0.01$ (4)  & $-0.07 \pm 0.06$ (4)  & $-0.31 \pm 0.01$ (4)  \\
Mg & $-0.26 \pm 0.05$ (10) & $-0.35 \pm 0.04$ (10) & $-0.26 \pm 0.06$ (10) & $-0.40 \pm 0.05$ (10) \\
Al & $-0.01 \pm 0.07$ (4)  & $-0.01 \pm 0.07$ (4)  & $ 0.01 \pm 0.08$ (4)  & $-0.07 \pm 0.08$ (4)  \\
Si & $-0.15 \pm 0.04$ (12) & $-0.16 \pm 0.03$ (12) & $-0.14 \pm 0.04$ (12) & $-0.17 \pm 0.04$ (12) \\
P  & $ 0.04 \pm 0.01$ (2)  & $-0.05 \pm 0.01$ (2)  & $ 0.12 \pm 0.01$ (2)  & $ 0.01 \pm 0.01$ (2)  \\
S  & $ 0.02 \pm 0.07$ (6)  & $-0.03 \pm 0.06$ (6)  & $ 0.09 \pm 0.07$ (6)  & $ 0.05 \pm 0.06$ (6)  \\
K  & $-0.19 \pm 0.05$ (2)  & $-0.17 \pm 0.02$ (2)  & $-0.16 \pm 0.05$ (2)  & $-0.20 \pm 0.03$ (2)  \\
Ca & $-0.07 \pm 0.03$ (9)  & $-0.12 \pm 0.03$ (9)  & $-0.05 \pm 0.03$ (9)  & $-0.17 \pm 0.03$ (9)  \\
Sc & $ 0.28 \pm \ldots$ (1)& $ 0.07 \pm 0.15$ (2)  & $ 0.35 \pm \ldots$ (1)& $-0.02 \pm 0.15$ (2)  \\
Ti & $-0.14 \pm 0.03$ (6)  & $-0.15 \pm 0.07$ (6)  & $-0.08 \pm 0.03$ (6)  & $-0.23 \pm 0.07$ (6)  \\
Cr & $-0.28 \pm 0.04$ (3)  & $-0.22 \pm 0.05$ (3)  & $-0.24 \pm 0.04$ (3)  & $-0.26 \pm 0.04$ (3)  \\
Co & $-0.24 \pm \ldots$ (1)& $-0.77 \pm \ldots$ (1)& $-0.18 \pm \ldots$ (1)& $-0.80 \pm \ldots$ (1)\\
Ni & $-0.18 \pm 0.05$ (5)  & $-0.17 \pm 0.03$ (5)  & $-0.14 \pm 0.05$ (5)  & $-0.18 \pm 0.03$ (5)  \\
Yb & $ 0.14 \pm \ldots$ (1)& $ 0.09 \pm \ldots$ (1)& $ 0.26 \pm \ldots$ (1)& $ 0.12 \pm \ldots$ (1)\\
\hline
\end{tabular}
\end{table}

\begin{table}[!ht]
\setlength{\tabcolsep}{10pt}
\centering
\small 
\renewcommand{\arraystretch}{0.99}
\caption{
Element-by-element abundance differences $\Delta$[X/H] (primary minus secondary) and their associated uncertainties for each wide binary system.
}
\label{tab:delta_abund}
\begin{tabular}{crrrr}
\hline\hline
\multirow{2}{*}{$\Delta$[X/H]} & {HD~20782~~~~} & {WASP-160~A~~~~} & {K2-54~~~~~~~~} & {WASP-127~~~~~~~}       \\
                               & {$-$ HD~20781} & {$-$ WASP-160~B} & {$-$ K2-54~B}   & {\tiny $-$ TYC~4916-897-1}  \\ 
\hline
C  & $-0.181 \pm 0.02$   & $-0.110 \pm 0.02$   & $-0.003 \pm 0.03$   & $0.125  \pm 0.05$   \\
N  & $0.057  \pm 0.05$   & $0.236  \pm 0.03$   & $-0.098 \pm 0.07$   & $0.245  \pm 0.09$   \\
O  & $0.112  \pm 0.10$   & $0.234  \pm 0.05$   & $-0.007 \pm 0.02$   & $0.158  \pm 0.11$   \\
Na & $-0.114 \pm 0.02$   & $-0.123 \pm 0.01$   & $0.013  \pm 0.07$   & $0.167  \pm 0.06$   \\
Mg & $-0.058 \pm 0.05$   & $0.000  \pm 0.06$   & $-0.078 \pm 0.04$   & $0.088  \pm 0.06$   \\
Al & $-0.098 \pm 0.11$   & $-0.041 \pm 0.10$   & $0.012  \pm 0.10$   & $0.005  \pm 0.10$   \\
Si & $-0.040 \pm 0.03$   & $-0.008 \pm 0.04$   & $-0.001 \pm 0.03$   & $0.008  \pm 0.05$   \\
P  & \ldots              & $-0.370 \pm \ldots$ & \ldots              & $0.090  \pm 0.02$   \\
S  & $-0.061 \pm 0.04$   & $-0.028 \pm 0.06$   & \ldots              & $0.049  \pm 0.09$   \\
K  & $0.015  \pm 0.01$   & $-0.060 \pm 0.03$   & $-0.005 \pm 0.03$   & $-0.025 \pm 0.06$   \\
Ca & $-0.016 \pm 0.02$   & $-0.038 \pm 0.04$   & $0.010  \pm 0.03$   & $0.047  \pm 0.05$   \\
Sc & $-0.075 \pm 0.02$   & $0.250  \pm \ldots$ & $0.085  \pm 0.07$   & $0.210  \pm 0.15$   \\
Ti & $-0.142 \pm 0.04$   & $-0.017 \pm 0.06$   & $-0.011 \pm 0.05$   & $0.004  \pm 0.08$   \\
Cr & $0.010  \pm 0.02$   & $-0.027 \pm 0.07$   & $0.020  \pm 0.05$   & $-0.060 \pm 0.06$   \\
Fe & $-0.104 \pm 0.03$   & $-0.021 \pm 0.03$   & $-0.003 \pm 0.03$   & $0.049  \pm 0.04$   \\
Co & $-0.190 \pm \ldots$ & $-0.210 \pm \ldots$ & $-0.010 \pm \ldots$ & $0.530  \pm \ldots$ \\
Ni & $-0.056 \pm 0.03$   & $-0.018 \pm 0.02$   & $-0.017 \pm 0.02$   & $-0.014 \pm 0.06$   \\
Yb & $-0.090 \pm \ldots$ & $-0.480 \pm \ldots$ & \ldots              & $0.050  \pm \ldots$ \\
\hline
\end{tabular}
\end{table}


\clearpage
\nolinenumbers

\section{Robustness checks for the $T_{\rm cond}$ slope analysis} \label{app:robust}
To assess the sensitivity of the inferred $T_{\rm cond}$ trends to methodological choices, we carried out two additional tests.
First, we repeated the same Monte Carlo analysis using the grid-fitting atmospheric parameter set instead of the preferred photometric parameter set.
Second, we repeated the analysis, including all measured elements, rather than restricting the fit to the 13 reliable elements adopted in the primary analysis.

Using the grid-fitting parameter set, the slopes of the differential abundances with $T_{\rm cond}$ are $-0.49 \pm 0.32$, $-1.01 \pm 0.24$, $+0.22 \pm 0.27$, and $-1.36 \pm 0.41$ $\times 10^{-4}$~dex~K$^{-1}$ for HD~20782/HD~20781, WASP-160~A/WASP-160~B, K2-54/K2-54~B, and WASP-127/TYC~4916-897-1, respectively.
These values are consistent within $1\sigma$ with those obtained using the photometric parameter set (see Figure~\ref{fig:T_cond}).
This comparison indicates that, although the abundances of individual elements can be derived somewhat differently depending on the adopted atmospheric parameter set, the overall comparisons between the binary components remain qualitatively stable when the analysis is performed in a homogeneous and internally consistent manner.

We also repeated the same analysis, including all measured elements.
In the primary analysis, we excluded five elements measured from fewer than three lines, namely P, K, Sc, Co, and Yb, in order to reduce the possibility that less precise measurements could introduce spurious trends.
The threshold of at least three lines was adopted as a practical criterion to ensure a minimum level of statistical reliability, since measurements based on only one or two lines are more vulnerable to intrinsic bias.
In addition, the absorption features of these elements are generally weaker and less secure than those of the other measured elements. 
A similar tendency is also seen in the comparison in Section~\ref{sec:sub:chem}, where the largest discrepancies with literature values are associated with elements measured from only a few spectral lines, such as Sc and Co.
For abundances derived from only one line, we adopted a measurement uncertainty of 0.1~dex in the Monte Carlo analysis.
Although the inclusion of these less secure measurements slightly affects the fitted slopes, the overall trends remain qualitatively consistent with those derived from the reliable element set.
The largest change is found for WASP-127/TYC~4916-897-1, for which the slope changes from $-1.12$ to $-0.56$ $\times 10^{-4}$~dex~K$^{-1}$, while the slopes of the other systems change only by about $0.1$ to $0.2$ $\times 10^{-4}$~dex~K$^{-1}$.


\end{appendix}


\end{document}